\documentstyle[twoside]{article}

\catcode`\@=11
\long\def\@makefntext#1{
\protect\noindent \hbox to 3.2pt {\hskip-.9pt
$^{{\eightrm\@thefnmark}}$\hfil}#1\hfill}               

\def\@makefnmark{\hbox to 0pt{$^{\@thefnmark}$\hss}}    

\def\ps@myheadings{\let\@mkboth\@gobbletwo
\def\@oddhead{\hbox{}
\rightmark\hfil\eightrm\thepage}
\def\@oddfoot{}\def\@evenhead{\eightrm\thepage\hfil
\leftmark\hbox{}}\def\@evenfoot{}
\def\sectionmark##1{}\def\subsectionmark##1{}}

\oddsidemargin=\evensidemargin
\newcounter{sectionc}\newcounter{subsectionc}\newcounter{subsubsectionc}
\renewcommand{\section}[1] {\vspace{12pt}\addtocounter{sectionc}{1}
\setcounter{subsectionc}{0}\setcounter{subsubsectionc}{0}\noindent
        {\tenbf\thesectionc. #1}\par\vspace{5pt}}
\renewcommand{\subsection}[1] {\vspace{12pt}\addtocounter{subsectionc}{1}
        \setcounter{subsubsectionc}{0}\noindent
        {\bf\thesectionc.\thesubsectionc. {\kern1pt \bfit #1}}\par\vspace{5pt}}
\renewcommand{\subsubsection}[1] {\vspace{12pt}\addtocounter{subsubsectionc}{1}
        \noindent{\tenrm\thesectionc.\thesubsectionc.\thesubsubsectionc.
        {\kern1pt \tenit #1}}\par\vspace{5pt}}
\newcommand{\nonumsection}[1] {\vspace{12pt}\noindent{\tenbf #1}
        \par\vspace{5pt}}

\newcounter{appendixc}
\newcounter{subappendixc}[appendixc]
\newcounter{subsubappendixc}[subappendixc]
\renewcommand{\thesubappendixc}{\Alph{appendixc}.\arabic{subappendixc}}
\renewcommand{\thesubsubappendixc}
        {\Alph{appendixc}.\arabic{subappendixc}.\arabic{subsubappendixc}}

\renewcommand{\appendix}[1] {\vspace{12pt}
        \refstepcounter{appendixc}
        \setcounter{figure}{0}
        \setcounter{table}{0}
        \setcounter{lemma}{0}
        \setcounter{theorem}{0}
        \setcounter{corollary}{0}
        \setcounter{definition}{0}
        \setcounter{equation}{0}
        \renewcommand{\thefigure}{\Alph{appendixc}.\arabic{figure}}
        \renewcommand{\thetable}{\Alph{appendixc}.\arabic{table}}
        \renewcommand{\theappendixc}{\Alph{appendixc}}
        \renewcommand{\thelemma}{\Alph{appendixc}.\arabic{lemma}}
        \renewcommand{\thetheorem}{\Alph{appendixc}.\arabic{theorem}}
        \renewcommand{\thedefinition}{\Alph{appendixc}.\arabic{definition}}
        \renewcommand{\thecorollary}{\Alph{appendixc}.\arabic{corollary}}
        \renewcommand{\theequation}{\Alph{appendixc}.\arabic{equation}}
        \noindent{\tenbf Appendix \theappendixc #1}\par\vspace{5pt}}
\newcommand{\subappendix}[1] {\vspace{12pt}
        \refstepcounter{subappendixc}
        \noindent{\bf Appendix \thesubappendixc. {\kern1pt \bfit #1}}
        \par\vspace{5pt}}
\newcommand{\subsubappendix}[1] {\vspace{12pt}
        \refstepcounter{subsubappendixc}
        \noindent{\rm Appendix \thesubsubappendixc. {\kern1pt \tenit #1}}
        \par\vspace{5pt}}

\newcommand{\textlineskip}{\baselineskip=13pt}
\newcommand{\smalllineskip}{\baselineskip=10pt}

\def\eightcirc{
\begin{picture}(0,0)
\put(4.4,1.8){\circle{6.5}}
\end{picture}}
\def\eightcopyright{\eightcirc\kern2.7pt\hbox{\eightrm c}}

\def\abstracts#1#2#3{{
        \centering{\begin{minipage}{4.5in}\baselineskip=10pt\footnotesize
        \parindent=0pt #1\par
        \parindent=15pt #2\par
        \parindent=15pt #3
        \end{minipage}}\par}}

\renewenvironment{thebibliography}[1]
        {\frenchspacing
         \ninerm\baselineskip=11pt
         \begin{list}{\arabic{enumi}.}
        {\usecounter{enumi}\setlength{\parsep}{0pt}
         \setlength{\leftmargin 12.7pt}{\rightmargin 0pt} 
         \setlength{\itemsep}{0pt} \settowidth
        {\labelwidth}{#1.}\sloppy}}{\end{list}}

\newcounter{itemlistc}
\newcounter{romanlistc}
\newcounter{alphlistc}
\newcounter{arabiclistc}

\newcommand{\fcaption}[1]{
        \refstepcounter{figure}
        \setbox\@tempboxa = \hbox{\footnotesize Fig.~\thefigure. #1}
        \ifdim \wd\@tempboxa > 5in
           {\begin{center}
        \parbox{5in}{\footnotesize\smalllineskip Fig.~\thefigure. #1}
            \end{center}}
        \else
             {\begin{center}
             {\footnotesize Fig.~\thefigure. #1}
              \end{center}}
        \fi}

\newcommand{\tcaption}[1]{
        \refstepcounter{table}
        \setbox\@tempboxa = \hbox{\footnotesize Table~\thetable. #1}
        \ifdim \wd\@tempboxa > 5in
           {\begin{center}
        \parbox{5in}{\footnotesize\smalllineskip Table~\thetable. #1}
            \end{center}}
        \else
             {\begin{center}
             {\footnotesize Table~\thetable. #1}
              \end{center}}
        \fi}

\def\@citex[#1]#2{\if@filesw\immediate\write\@auxout
        {\string\citation{#2}}\fi
\def\@citea{}\@cite{\@for\@citeb:=#2\do
        {\@citea\def\@citea{,}\@ifundefined
        {b@\@citeb}{{\bf ?}\@warning
        {Citation `\@citeb' on page \thepage \space undefined}}
        {\csname b@\@citeb\endcsname}}}{#1}}

\newif\if@cghi

\def\pmb#1{\setbox0=\hbox{#1}
        \kern-.025em\copy0\kern-\wd0
        \kern.05em\copy0\kern-\wd0
        \kern-.025em\raise.0433em\box0}

\def\fnt#1#2{\footnotetext{\kern-.3em
        {$^{\mbox{\scriptsize #1}}$}{#2}}}

\def\fpage#1{\begingroup
\voffset=.3in
\thispagestyle{empty}\begin{table}[b]\centerline{\footnotesize #1}
        \end{table}\endgroup}

\def\runninghead#1#2{\pagestyle{myheadings}
\markboth{{\protect\footnotesize\it{\quad #1}}\hfill}
{\hfill{\protect\footnotesize\it{#2\quad}}}}
\font\tenrm=cmr10
\font\tenit=cmti10
\font\tenbf=cmbx10
\font\bfit=cmbxti10 at 10pt
\font\ninerm=cmr9

\font\eightrm=cmr8

\def\qed{\hbox{${\vcenter{\vbox{                        
   \hrule height 0.4pt\hbox{\vrule width 0.4pt height 6pt
   \kern5pt\vrule width 0.4pt}\hrule height 0.4pt}}}$}}

\begin{document}

\input{psfig.tex}

\def\VEV#1{{\left\langle #1 \right\rangle}}
\def\lesssim{\mathrel{\rlap{\lower4pt\hbox{\hskip1pt$\sim$}}
    \raise1pt\hbox{$<$}}}         
\def\gtrsim{\mathrel{\rlap{\lower4pt\hbox{\hskip1pt$\sim$}}
    \raise1pt\hbox{$>$}}}         
\def\hatn{{\bf \hat n}}

\runninghead{Theory of Cosmic Microwave Background
Polarization}{Cabella and Kamionkowski}

\normalsize\textlineskip
\thispagestyle{empty}
\setcounter{page}{1}


\vspace*{0.88truein}

\fpage{1}
\centerline{\bf THEORY OF COSMIC MICROWAVE BACKGROUND
POLARIZATION\footnote{Lectures given at the 2003 Villa
Mondragone School of Gravitation and Cosmology: ``The Polarization
of the Cosmic Microwave Background,'' Rome, Italy, September
6--11, 2003.}}
\vspace*{0.37truein}
\centerline{\footnotesize PAOLO CABELLA\footnote{Paolo.Cabella@roma2.infn.it}}
\vspace*{0.015truein}
\centerline{\footnotesize\it Dipartimento di Fisica,
Universit\`a Tor Vergata, Roma I-00133, Italy}
\vspace*{10pt}
\centerline{\footnotesize MARC KAMIONKOWSKI\footnote{kamion@tapir.caltech.edu}}
\vspace*{0.015truein}
\centerline{\footnotesize\it California Institute of Technology, Mail Code
130-33, Pasadena, CA~~91125, USA}

\vspace*{0.21truein}
\abstracts{These lectures introduce some of the basic theory of
cosmic microwave background (CMB) polarization with the primary
aim of developing the theory of CMB polarization from
inflationary gravitational waves, as well as some of the related
theory of weak gravitational lensing (cosmic shear) of CMB
polarization.  We begin with production of polarization by Thomson
scattering.  We then discuss tensor-harmonic analysis (the
``grad-curl'' or ``E-B'' decomposition) on a flat and full sky
in some detail.  The Boltzmann/Einstein equations required to
predict the CMB temperature/polarization pattern due to
primordial gravitational waves are derived.  We show that
gravitational waves produce a curl component of the CMB
polarization while density perturbations (at linear order) do
not. We then show how cosmic shear induces a curl component from
a curl-free surface of last scattering.  We describe, though in
less detail, how higher-order correlations can be used to
subtract the cosmic-shear--induced curl.  Several exercises are
provided.}{}{}


\vspace*{1pt}\textlineskip      

\section{Introduction}

Just five years ago, the small-scale structure of the CMB was
hidden behind a veil of experimental limitation.  That is no
longer the case.  We now know empirically that the CMB has a
wealth of detailed information in its temperature
\cite{cmbexperiments} and polarization pattern
\cite{polarizationdetections}.  We are now also confident in our
theoretical understanding of these fluctuations: they are
produced by a nearly scale-invariant spectrum of primordial
density perturbations.  With a solid theoretical foundation,
we are poised to move considerably further with future more
precise CMB experiments.  Prospects for new advances include an
even more detailed picture of the physics at the surface of last
scattering, the physics of inflation, gravitational-wave
backgrounds, the distribution of mass in the
intermediate-redshift and current Universe, and perhaps the
behavior of the dark energy.

These lectures will {\it not} discuss all these wonderful
possibilities.  Instead, we focus here on detailing
some of the basics of the theory of CMB polarization with
the aim primarily of understanding the polarization due to
inflationary gravitational waves, as well as the related effects
of weak gravitational lensing (``cosmic shear'') on CMB
polarization.
There is now a vast literature on CMB theory, including a
number of reviews \cite{reviews} and even a few
books \cite{books}.  Thus,
experts are unlikely to find anything new in here.  However,
some students may find that the discussion and level of
detail for the specific subjects on which we focus
provide a useful complement to these other more comprehensive
sources.  In particular, we discuss tensor harmonics (the
``curl-grad'' or ``E-B'' decomposition) on both a
flat and a full sky in detail, and we provide a detailed
justification, from scratch, of the statement that
gravitational waves produce a curl component in the CMB while
density perturbations do not.  We show in detail that cosmic
shear induces a curl component in the CMB polarization, but our
discussion of how this shear may be subtracted with higher-order
correlations is a bit sketchier.  We do not deal with any
of the many important issues in data analysis.

The plan of the lectures is as follows:  We begin in Section 2
by reviewing how Thomson scattering produces polarization.
Section 3 develops tensor harmonic analysis on a flat sky and
Section 4 deals briefly with flat-sky correlation functions.
Section 5 deals with tensor harmonic analysis on the full sky
and Section 6 with the correlation
functions.  Section 7 provides some comments, and in Section 8
we discuss some of the effects of cosmic shear on the
polarization pattern.  A few exercises are sprinkled throughout
to help conscientious students develop their facility with
subject matter.  We reference primarily the most important
original papers on relevant subjects and/or those that were
particularly useful to us in the preparation of these lectures,
but our referencing does not provide a full guide to the
literature.  For that we refer the reader to the more
comprehensive reviews \cite{reviews}.

\section{Polarization of CMB: Theory}

\begin{figure}[tbp]
\centerline{\psfig{file=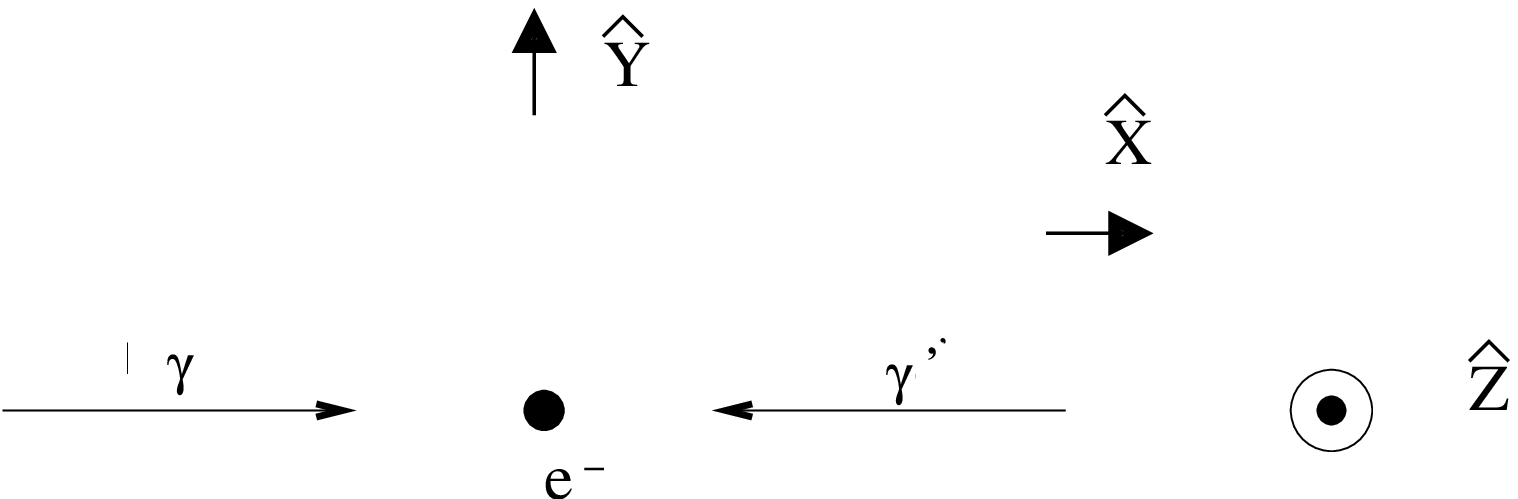,width=12cm}}
\bigskip
\fcaption{Radiation incident on an electron from the $\pm {\bf \hat x}$
directions scatters into the ${\bf \hat z}$ direction by shaking
the electron in the $\pm {\bf \hat y}$ direction.}
\label{fig:scatter}
\end{figure}

Let us begin by understanding in the most basic terms why the
CMB should be polarized \cite{rees}. Consider scattering of an
electromagnetic wave propagating in the $\pm\hat{x}$ direction
in the plane of the page that then scatters to the observer
(perpendicular to the page) by $90^\circ$; the scatterer is a
free electron (Fig. \ref{fig:scatter}).
If the incident radiation is linearly polarized with an
$\vec{E}$ vector in the $\hat{z}$ direction, then the electron
oscillates in the $\hat{z}$ direction and emits no dipole radiation
toward the observer. But if the incident radiation is polarized
in the $\hat{y}$ direction, the electron oscillates in the $\hat{y}$
direction and the dipole radiation the observer sees is linearly
polarized in the $\pm\hat{y}$ direction. If the incident
radiation is unpolarized then only the component with
$\vec{E}||\hat{y}$ is scattered toward the observer, and the
resulting scattered radiation is polarized in the $\hat{y}$
direction.  Likewise if the incident radiation comes only from the
$\pm\hat{y}$ direction, then the observed scattered radiation
will be polarized in the  $\pm\hat{x}$ direction.

\begin{figure}[tbp]
\centerline{\psfig{file=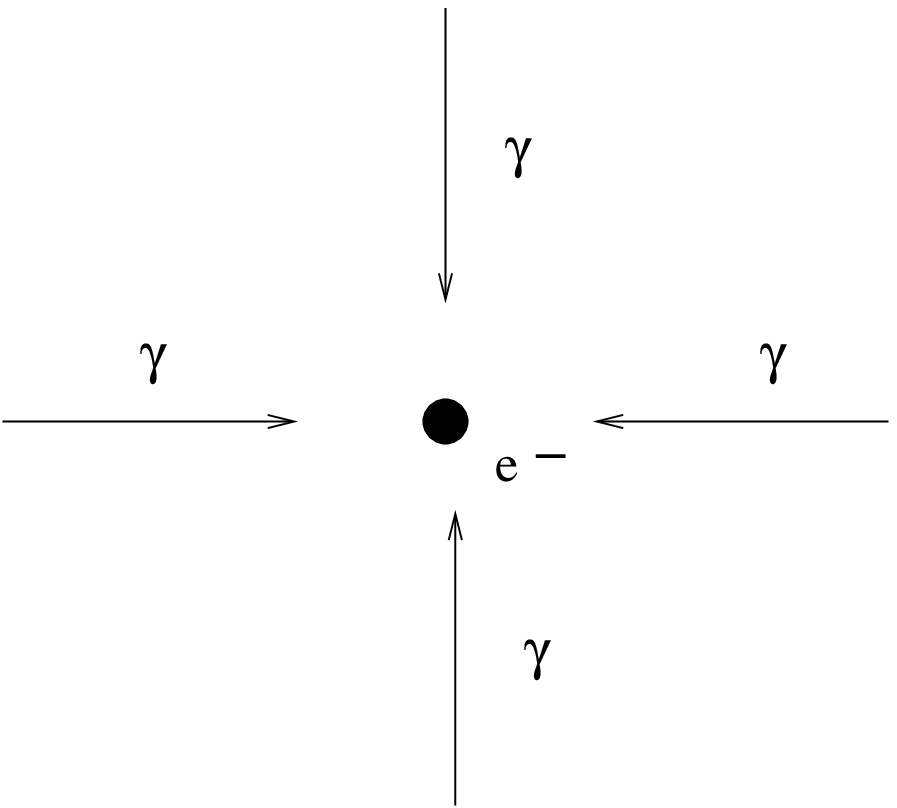,width=10cm}}
\bigskip
\fcaption{Radiation incident on the electron with equal
intensities in the $\pm {\bf \hat x}$ and $\pm {\bf \hat y}$
directions will be unpolarized when scattered to the ${\bf \hat
z}$ direction.}
\label{fig:isorad}
\end{figure}

Now if we have radiation incident from the $\pm\hat{x}$ and
$\pm\hat{y}$ directions with equal intensities (see
Fig. \ref{fig:isorad}), then the
polarizations in the $\pm\hat{x}$ and $\pm\hat{y}$ directions
will cancel, and the scattered radiation will be unpolarized.
But if the intensity in the $\hat{x}$ direction slightly exceeds
that in the $\hat{y}$ direction, then the scattered radiation
will be slightly polarized in the $\hat{y}$ direction.

\begin{figure}[tbp]
\centerline{\psfig{file=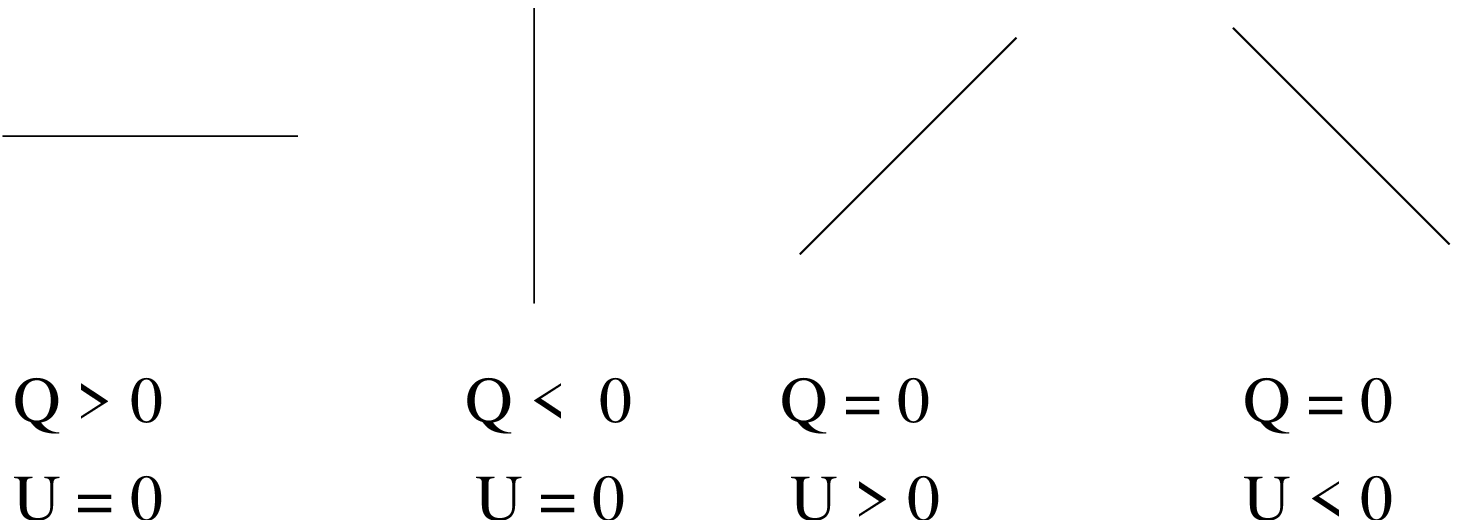,width=12cm}}
\bigskip
\fcaption{The linear-polarization states described various
combinations of Stokes parameters $Q$ and $U$.}
\label{fig:polrot}
\end{figure}

To proceed further we must now introduce the Stokes parameters
$Q$ and $U$ for linear polarization. Most
generally, a monochromatic electromagnetic wave propagating in
the $\hat{z}$ direction has an electric-field vector,
\begin{equation}
      E_x = a_x\cos(\omega t-\xi_x); \qquad E_y =
      a_y\cos(\omega t-\xi_y).
\end{equation}
The Stokes parameters are then the intensity,
\begin{equation}
     I= a^2_x+a^2_y,
\end{equation}
linear-polarization parameters,
\begin{equation}
     Q=a^2_x-a^2_y, \qquad U=2a_xa_y\cos(\xi_x-\xi_y),
\end{equation}
and circular-polarization parameter,
\begin{equation}
     V=2a_xa_y\sin(\xi_x-\xi_y).
\end{equation}
The last Stokes parameter $V$ vanishes as Thomson scattering
induces no circular polarization (but see
Ref. \cite{circular}). The parameter $Q$ quantifies
the polarization in the $x$-$y$ directions while $U$ quantifies it
along axes rotated by $45^\circ$.  Fig. \ref{fig:polrot}
illustrates the linear polarization described by various
combinations of $Q$ and $U$.

If we rotate the $x$-$y$ axes by an angle $\alpha$ about
the line of sight $\hat{z}$, then the new $x'$-$y'$ coordinates
are
\begin{equation}
 \left(
\begin{array}{c}
x'\\
y'\\
\end{array}
\right)
=
\left(
\begin{array}{cc}
\cos\alpha & \sin\alpha\\
-\sin\alpha & \cos\alpha \\
\end{array}
\right)
\left(
\begin{array}{c}
x\\
y\\
\end{array}
\right),
\end{equation}
but the Stokes parameters $(Q,U)$ transform as \cite{Chandrasekhar}
\begin{equation}
\left(
\begin{array}{c}
Q'\\
U'\\
\end{array}
\right)
=
\left(
\begin{array}{cc}
\cos2\alpha & \sin2\alpha\\
-\sin2\alpha & \cos2\alpha \\
\end{array}
\right)
\left(
\begin{array}{c}
Q\\
U\\
\end{array}
\right).
\end{equation}
Formally, $(Q,U)$ are two quantities that under a coordinate
transformation,
\begin{equation}
     x'_i = A^k_ix_k,
\end{equation}
transform as
\begin{equation}
\label{eq.tenstrans}
     P'_{ij}= A^k_iA^l_jP_{kl}.
\end{equation}
More explicitly, $(Q,U)$ are the components of a symmetric
trace-free $2\times 2$ tensor,
\begin{equation}
\left(
\begin{array}{cc}
Q& U\\
U & -Q\\
\end{array}
\right )
\Rightarrow
\left(
\begin{array}{cc}
\cos\alpha & \sin\alpha\\
-\sin\alpha & \cos\alpha \\
\end{array}
\right)
\left(
\begin{array}{cc}
Q& U\\
U & -Q\\
\end{array}
\right )
\left(
\begin{array}{cc}
\cos\alpha & -\sin\alpha\\
\sin\alpha & \cos\alpha \\
\end{array}
\right),
\end{equation}
or alternatively and equivalently, a  spin-2 field.

We now return to the generation of polarization by Thomson
scattering.  Generalizing our earlier heuristic results, the
magnitude of the polarization of the scattered
radiation is proportional to the magnitude of the quadrupole of
the radiation incident on the scattering electron, and the
orientation of the polarization is determined by the orientation
of the quadrupole.  More precisely, from the angular dependence
\cite{jackson},
\begin{equation}
     \frac{d\sigma}{d\Omega}=\frac{3\sigma_T}
     {8\pi}|\hat{\epsilon}'\cdot\hat{\epsilon}|^2,
\end{equation}
of Thomson scattering, where $\sigma_T$ is the Thomson cross
section and $\hat{\epsilon}'$   and $\hat{\epsilon}$ are the
polarization of the incident and scattered radiation,
respectively, it can be shown that the polarization of radiation
scattered from an electron cloud of optical depth $\tau\ll 1$
into the $\hat{z}$ direction is
\begin{equation}
     Q-iU=\sqrt{\frac{3}{40\pi}}\tau a_{22}.
\label{eqn:QiU}
\end{equation}
Here, $a_{22}$ is the radiation quadrupole moment incident on the
electron cloud. More precisely, $a_{22}$ is the coefficient of
the spherical harmonic $Y_{22}(\theta,\phi)$ in a
spherical-harmonic expansion of the incident-radiation intensity
in a coordinate system in which the line of sight is the
$\hat{z}$ direction, $\hat x$ and $\hat z$ are in the scattering
plane, and $Q$ and $U$ are measured with respect to the $x$ and
$y$ axes.  (See Ref. \cite{newastron} for expressions for more
general orientations.)

\medskip
\noindent {\sl Exercise 1. Verify Eq. (\ref{eqn:QiU}).}
\medskip

\begin{figure}[tbp]
\centerline{\psfig{file=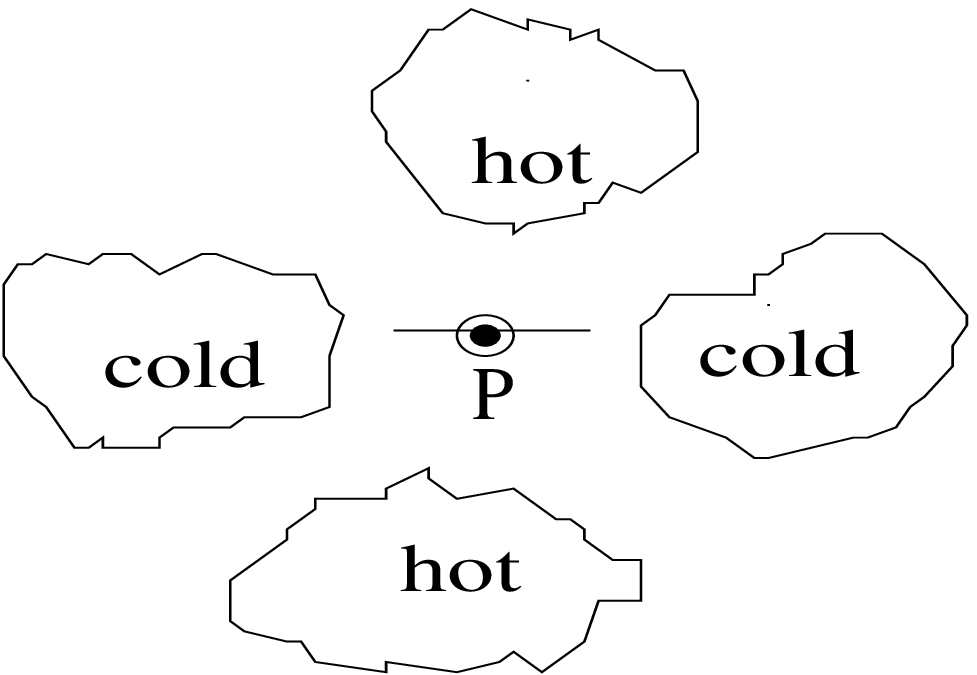,width=12cm}}
\bigskip
\fcaption{The radiation scattered from the center of this
quadrupolar temperature pattern will be polarized as shown.}
\label{fig:hotcoldspots}
\end{figure}

Finally (!) we can see why the CMB must be polarized. We see
angular variations in the temperature of the CMB implying
temperature (and thus intensity) inhomogeneities at the CMB
surface of last scattering. Thus, most generally, a given scatterer
at the last-scattering surface will see an anisotropic
distribution of incident radiation leading to polarized
scattered radiation. For example, the radiation in the center of
Fig. \ref{fig:hotcoldspots}  will be polarized in the
direction of the two cold spots in the temperature field.

\section{Harmonic analysis for $Q$ and $U$ on a flat sky}

According to inflation (as well as other structure-formation
theories), the temperature pattern on the sky is a single
realization of a random field. The simplest (and for a Gaussian
field only non-trivial) statistic to describe the temperature
pattern is the power spectrum $C^{\rm TT}_l$.

Suppose that we now have in addition to the temperature pattern the
polarization, quantified by Stokes parameters $Q(\hat{n})$ and
$U(\hat{n})$ measured as a function of position
$\hat{n}=(\theta,\phi)$ on the sky. Again, since $Q$ and $U$ are
components of a symmetric trace-free (STF) $2\times2$ tensor, they
depend on the coordinate system we choose. To discuss
physics, we will thus want to find a
coordinate-system--independent representation of this tensor
field.  Later, we will need to do this on the full sky, but as a
warmup, we will first do the simpler case of a flat sky (which
also serves as a good approximation to a small region of the
sky).

Before moving further, it should be noted that the
formalism we are about to develop also applies to cosmic-shear
(weak gravitational lensing) maps, where the ellipticity
parameters $\epsilon_+$ and $\epsilon_\times$ are, like $Q$ and
$U$, components of a STF $2\times2$ tensor \cite{weaklens}.

Once we have measured the polarization, $Q(\vec{\theta})$ and
$U(\vec{\theta})$, as a function of position
$\vec{\theta}=(\theta_x,\theta_y)$ on a flat region of sky, we
have measured the polarization tensor field,
\begin{equation}
P_{ab} = \frac{1}{2}\left(
\begin{array}{cc}
Q(\vec{\theta})& U(\vec{\theta})\\
U(\vec{\theta}) & -Q(\vec{\theta})\\
\end{array}
\right ),
\label{eqn:QUmatrix}
\end{equation}
which is sometimes written as a vector, $P_a(Q,U)=P(Q,U)$,
although this is a dangerous thing to do as $Q$ and $U$ do not
transform as the components of a vector. The polarization is
also sometimes written as a complex number,
\begin{equation}
     P=|P|e^{2i\alpha}=(Q^2+U^2)^{1/2}e^{2i\alpha}=Q+iU,
\end{equation}
where $|P|=(Q^2+U^2)^{1/2}$ is the magnitude of the polarization and
$\alpha=(1/2)\arctan(U/Q)$ is its orientation relative to the
$x$ axis.

We now define gradient (`G', also known as E) and curl (`C',
also known as B) components of the tensor field that are
independent of the orientation of the $x$-$y$ axes as follows:
\begin{equation}
\label{eq.nabla2}
     \nabla^2 P_{\rm G}=\partial_a \partial_bP_{ab}\:\:\:;\;\:\:
     \nabla^2 P_{\rm C}
     = \epsilon_{ac}\partial_b\partial_cP_{ab},
\end{equation}
where $\epsilon_{ab}$ is the antisymmetric tensor. 

We can write more explicit expressions for these G and C
components in Fourier space.  Writing
\begin{equation}
     P_{ab}(\vec{\theta}) = \int\frac{d^2\vec{l}}{(2\pi)^2}
     \tilde{P}_{ab}(\vec{l}) e^{-i\vec{l}\cdot\vec{\theta}},
\end{equation}
\begin{equation}
\tilde{P}_{ab}(\vec{l})=\int
     d^2\vec{\theta}{P}_{ab}(\vec{\theta})e^{i\vec{l}\cdot\vec{\theta}},
\end{equation}
the Fourier components of $P_{\rm G}(\vec{\theta})$ and $P_{\rm
C}(\vec{\theta})$ are
\begin{eqnarray}
     \tilde{P}_{\rm G}(\vec{l}) & = & \frac{1}{2} \frac{(l^2_x-l^2_y)
     \tilde{Q}(\vec{l}) +2l_xl_y\tilde{U}(\vec{l})}{l^2_x+l^2_y},\\
     \tilde{P}_{\rm C}(\vec{l}) & = & \frac{1}{2}
     \frac{2l_xl_y\tilde{Q}
     (\vec{l})-(l^2_x-l^2_y)\tilde{U}(\vec{l})}{l^2_x+l^2_y}.
\label{eqn:GCFouriercomponents}
\end{eqnarray}

\medskip
\noindent {\sl Exercise 2. Verify these expressions for the
Fourier components $\tilde P_{\rm G}$ and $\tilde P_{\rm C}$, and verify
that they are invariant under a rotation of the
$\theta_x-\theta_y$ axes.}
\medskip

For a temperature map $T(\vec{\theta})$, the temperature
power spectrum $C_l^{\rm TT}$ is defined from
\begin{equation}
     \VEV{\tilde{T}(\vec{l})\tilde{T}(\vec{l'})} = (2\pi)^2
     \delta(\vec{l}+\vec{l'})C_l^{\rm TT},
\end{equation}
where the angle brackets denote an average over all
realizations. Likewise, the statistics of the polarization field
are determined by polarization power spectra $C_l^{\rm GG}$,
$C_l^{\rm CC}$, and $C_l^{\rm GC}$ defined by
\begin{eqnarray}
     \VEV{\tilde{P}_{\rm G}(\vec{l})\tilde{P}_{\rm G}(\vec{l}')} & = &
     (2\pi)^2\delta(\vec{l}+\vec{l}')C_l^{\rm GG},\\
     \VEV{\tilde{P}_{\rm C}(\vec{l})\tilde{P}_{\rm C}(\vec{l}')} & = & (2\pi)^2
     \delta(\vec{l}+\vec{l}')C_l^{\rm CC},\\ 
     \VEV{\tilde{P}_{\rm G}(\vec{l})\tilde{P}_{\rm C}(\vec{l}')}
     & = & (2\pi)^2 \delta(\vec{l}+\vec{l}')C_l^{\rm GC}.
\end{eqnarray}
If we also consider cross-correlation of the polarization with
temperature, then there are in total six power spectra,
\begin{equation}
     \VEV{\tilde{X}_1(\vec{l})\tilde{X}_2(\vec{l'})} = (2\pi)^2
     \delta(\vec{l}+\vec{l'})C_l^{X_1X_2},
\end{equation}
where $X_1, X_2 =\{ T, P_{\rm G}, P_{\rm C}\}$.

Now suppose we have a given temperature/polarization map and
then consider a parity inversion; e.g., consider a reflection
about the $x$-axis. Then 
\begin{equation}
     \theta_y\rightarrow-\theta_y,\quad Q\rightarrow Q,\quad U\rightarrow
     -U,\quad l_x\rightarrow l_x, \quad l_y\rightarrow -l_y.
\end{equation}
Also,
\begin{equation}
\tilde{T}(\vec{l})\rightarrow\tilde{T}(\vec{l}),\:\:\:
\tilde{P}_{\rm G}(\vec{l})\rightarrow\tilde{P}_{\rm G}(\vec{l}),\:\:\:
\tilde{P}_{\rm C}(\vec{l})\rightarrow -\tilde{P}_{\rm C}(\vec{l}).
\end{equation}
In other words, G and T are parity even, while C is parity
odd. Thus, if the physics that gives rise to T/P fluctuations
is parity conserving, we expect
\begin{equation}
     C_l^{\rm TC}=C_l^{\rm GC}=0,
     \:\:\:\:\:\:\:\mathrm{\emph{parity invariance}}
\end{equation}
in which case the statistics of the T/P  map is determined
entirely by the four power spectra, $ C_l^{\rm TT},
C_l^{\rm TG},C_l^{\rm GG}$, and $C_l^{\rm CC}$.

\section{Correlation functions}

\begin{figure}[tbp]
\centerline{\psfig{file=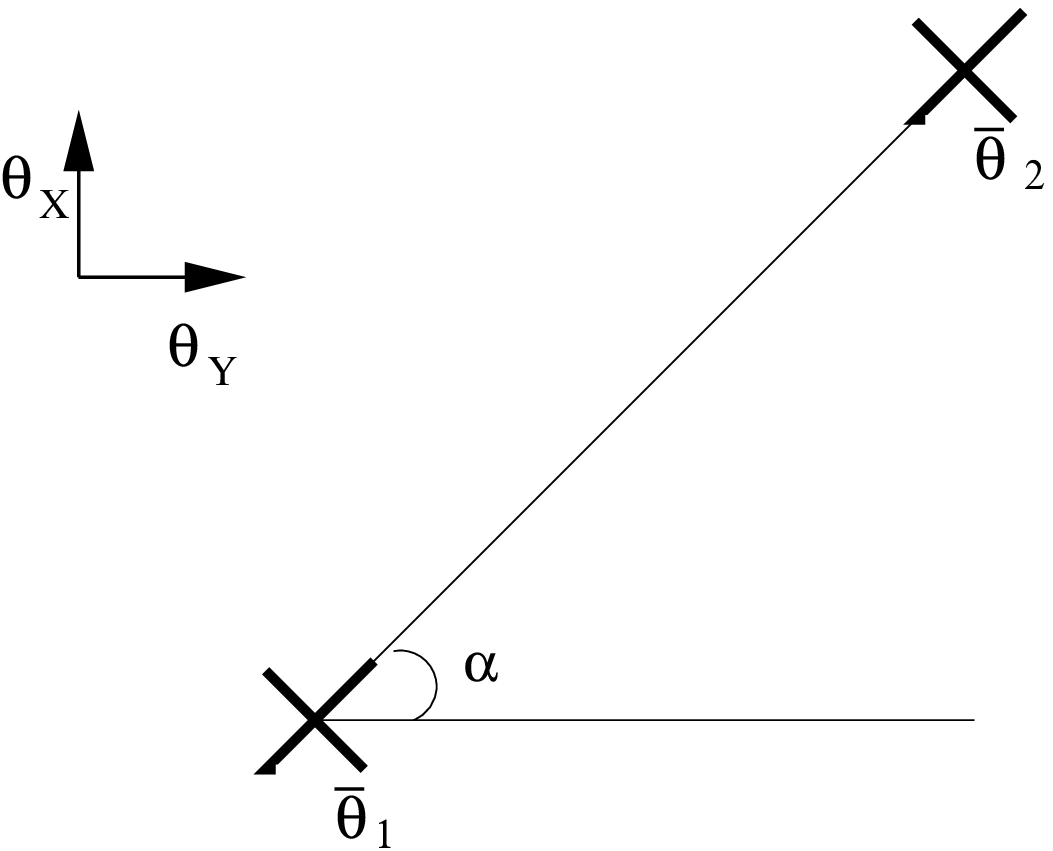,width=10cm}}
\fcaption{Correlation functions for polarization are best
defined for Stokes parameters $Q_r$ and $U_r$ measured with
respect to axes aligned with the line connecting the two points
being correlated.}
\label{fig:rotated}
\end{figure}

For a temperature map, the correlation between the temperature
$T(\vec{\theta}_1)$ and $T(\vec{\theta}_2)$ at any two points
is 
\begin{equation}
     \VEV{T(\vec{\theta}_1)T(\vec{\theta}_2)} =
     C_{\rm TT}\left(|\vec{\theta}_1-\vec{\theta}_2|\right),
\end{equation}
and the correlation function $C_{\rm TT}(\theta)$ depends only
on the distance $\theta$ between the two points.  The
same is not true for $Q$ and $U$. Since $Q$ and $U$ are not
rotational invariants, their correlations will depend also on
the orientation of the line connecting the two points. Instead,
one can define rotationally invariant correlation functions by
considering the components $Q_r$ and $U_r$ of the polarization
defined with respect to the line connecting the two points
(Fig. \ref{fig:rotated}).
We can then define the correlation functions,
\begin{eqnarray}
\VEV{Q_r(\vec{\theta}_1)Q_r(\vec{\theta}_2)}&=&
C_{\rm QQ}\left(|\vec{\theta}_1-\vec{\theta}_2|\right),\\
\VEV{U_r(\vec{\theta}_1)U_r(\vec{\theta}_2)}&=&
C_{\rm UU}\left(|\vec{\theta}_1-\vec{\theta}_2|\right),\\
\VEV{Q_r(\vec{\theta}_1)U_r(\vec{\theta}_2)}&=&
C_{\rm QU}\left(|\vec{\theta}_1-\vec{\theta}_2|\right).
\end{eqnarray}
Since $Q\rightarrow Q$ and $U\rightarrow -U$ under a parity
inversion, $C_{\rm QU}=0$ if parity is conserved.  If we
correlate the polarization with temperature, then there are
another three (TT,TQ,TU) correlation functions, one of which
(TU) vanishes if parity is conserved.

The correlation functions can be written in terms of the power
spectra,
\begin{eqnarray}
     C_{\rm QQ}(\theta)&+&C_{\rm UU}(\theta) = -\int_0^\infty\frac{l\,dl}{\pi}
     [C_l^{\rm GG}+C_l^{\rm CC}]J_0(l\theta),\\ 
     C_{\rm QQ}(\theta) & - & C_{\rm UU}(\theta) =
     -\int_0^\infty\frac{l\,dl}{\pi} [C_l^{\rm GG}-C_l^{\rm CC}]J_4(l\theta),
\end{eqnarray}
where $J_\nu(x)$ are Bessel functions.  These relations can also
be inverted to give power spectra in terms of correlation
functions. For derivations and details, see Refs. \cite{weaklens,kks}.

Finally, the correlation functions at zero lag give us the
mean-square polarization,
\begin{equation}
     \VEV{P^2}=\VEV{Q^2+U^2} = \int_0^\infty\frac{l\,dl}{2\pi}
     [C_l^{\rm GG}+C_l^{\rm CC}]=\VEV{P^2_{\rm G}}+\VEV{P^2_{\rm
     C}}. 
\end{equation}

\medskip
\noindent {\sl Exercise 3. Write an expression for the
temperature power spectrum $C_l^{\rm TT}$ in terms of the
temperature autocorrelation function $C^{\rm TT}(\theta)$.  Now
do the same for the gradient and curl power spectra in terms of
the QQ and UU autocorrelation functions.}
\medskip

\section{Harmonic analysis on the full sky}

If our maps extend beyond a small region of the sky, we will
have to come to terms with the fact that the sky
is a curved surface. Moreover, as we will see below, theoretical
predictions for the power spectra will require solution of
Boltzmann equations that require that we keep track of the
distribution function for photons moving in all directions.

We thus generalize the tensor Fourier analysis (G and C)
that we did above for STF $2\times 2$ tensors that live on the
2-sphere. Our discussion in this Section follows
Ref. \cite{kks}; a different but equivalent formalism is
presented in Ref. \cite{zstensor}.  In the usual spherical polar
coordinates $\theta,\phi$, the sphere has a metric,
\begin{equation}
g_{ab}=
\left(
\begin{array}{cc}
1 & 0\\
0 & \sin^2\theta\\
\end{array}
\right).
\label{metric}
\end{equation}
The polarization tensor $P_{ab}$ must be symmetric $P_{ab}=P_{ba}$ and
trace-free $g^{ab}P_{ab}=0 $, from which it follows that,
\begin{equation}
P_{ab}(\hat{n})= \frac{1}{2}
\left(
\begin{array}{cc}
Q(\hat{n}) & -U(\hat{n})\sin\theta\\
-U(\hat{n})\sin\theta & -Q(\hat{n})\sin^2\theta\\
\end{array}
\right).
\end{equation}
(The alert reader will note that the sign of $U$ has changed
here relative to Eq. (\ref{eqn:QUmatrix}).  This is
simply because for the full sky we are following the slightly
different conventions used in Ref. \cite{kks}.)
The factors of $\sin\theta$ also follow from the fact that the
coordinate basis $(\theta,\phi)$ is orthogonal but not
orthonormal.

Let us now review some of the rules of differential geometry on the
two-sphere.  Covariant derivatives of scalar, vector, and tensor
fields are 
\begin{equation}
S_{:a}=S_{,a}, \qquad
V^a{}_{:b}=V^a{}_{,b}+V^c\Gamma^a_{bc}, \qquad
T^{ab}{}_{:c}=T^{ab}{}_{,c}+T^{db}\Gamma^a_{cd}+T^{ad}\Gamma^b_{cd},
\end{equation}
and `:' denotes covariant derivative, $S_{,a}=(\partial S/\partial
x^\alpha)$, and the Christoffel symbols are
\begin{equation}
     \Gamma^a_{bc}=\frac{1}{2}g^{ad}(g_{db,c}+g_{dc,b}-g_{bc,d}) .
\end{equation}
Also,
\begin{equation}
     S^{:ab}{}_{ab}=\nabla^2\nabla^2S+R^{db}S_{:db} +
     {1\over2}\,R^{:d}S_{:d}, \quad  \nabla^2S\equiv S^{:a}{}_a
     \quad R_{ab}\equiv R^c{}_{acb}, \quad R\equiv R^a{}_a.
\end{equation}
Since the sphere has no boundary, we can integrate by parts,
\begin{equation}
     \oint d^2\hat{n} \sqrt{g} X^{ab} Y_{:ab}
     = -\oint d^2\hat{n} \sqrt{g} X^{ab}{}_{:a} Y_{:b}
     = \oint d^2\hat{n} \sqrt{g} X^{ab}{}_{:ba} Y,
\label{byparts}
\end{equation}
where $Y(\hat{n})$ is a scalar function and $\oint d^2\hat{n}$
denotes integration over the sphere. The antisymmetric tensor is
\begin{equation}
\epsilon_{ab}=\sqrt{g}
\left(
\begin{array}{cc}
 0 & 1 \\
-1 & 0
\end{array}
\right),
\end{equation}
and
\begin{equation}
\epsilon_{ca}\epsilon^c{}_b=g_{ab}=-\epsilon_{ac}\epsilon^c{}_b, \qquad
\epsilon_{ab}\epsilon_{cd}  =g_{ac}g_{bd}-g_{ad} g_{bc},         \qquad
\epsilon_{ab:c}=0.
\label{epsproperties}
\end{equation}
Also, $\epsilon^{ab}T_{ab}=0$ for a symmetric tensor, $T_{ab}=T_{ba}$.

For the two-sphere, the Riemann tensor is
\begin{equation}
              R_{abcd}={1\over2}\,R\,\epsilon_{ab}\epsilon_{cd}, \quad
                R_{ab}={1\over2}\,R\,g_{ab},                    \quad
 \epsilon^{ab}R_{abcd}=           R\,\epsilon_{cd},             \quad
 \epsilon^{ac}R_{abcd}={1\over2}\,R\,\epsilon_{bd} .
\end{equation}
For two STF tensors $M_{ab}, N_{ab}$ (i.e.
$g^{ab}M_{ab}=g^{ab}N_{ab}=\epsilon^{ab}M_{ab}=\epsilon^{ab}N_{ab}=0)$,
the second of Eqs. (\ref{epsproperties}) gives us
\begin{equation}
     M^{ab}N^{cd}\epsilon_{ac}\epsilon_{bd}=-M^{ab}N_{ab}.
\end{equation}
Also,
\begin{equation}
g=|g_{ab}|=\sin^2\theta, \qquad
\epsilon^a_{b}=
\left(
\begin{array}{cc}
 0 & \sin\theta \\
-\csc\theta & 0
\end{array}
\right),
\end{equation}
\begin{equation}
\Gamma^\theta_{\phi\phi}= -\sin\theta\cos\theta, \qquad
\Gamma^\phi_{\theta\phi}=\Gamma^{\phi}_{\phi\theta}=\cot\theta, \qquad
\end{equation}
and all other components vanish. Then, for a scalar function $Y(\hat{n})$,
\begin{eqnarray}
Y_{:\theta\theta} &=& Y_{,\theta\theta}, \nonumber\\
Y_{:\theta\phi} &=& Y_{,\theta\phi}
    -\cot\theta\, Y_{,\phi},\nonumber\\
Y_{:\phi\phi} &=& Y_{,\phi\phi}+\sin\theta\cos\theta Y,_{\theta}.
\end{eqnarray}
A symmetric rank-2 tensor $M_{ab}$ has `divergence',
\begin{eqnarray}
     M^{ab}{}_{:ab} = & & M^{\theta\theta}{}_{,\theta\theta} +
     2 M^{\theta\phi}{}_{,\theta\phi} +
     M^{\phi\phi}{}_{,\phi\phi}
      - \sin\theta\cos\theta M^{\phi\phi}{}_{,\theta}
     \nonumber\\
     & + &2 \cot\theta  M^{\theta\theta}{}_{,\theta}
     + 4 \cot\theta   M^{\theta\phi}{}_{,\phi}
     + (1-3\cos^2\theta) M^{\phi\phi} - M^{\theta\theta},
\label{MderivsG}
\end{eqnarray}
and `curl',
\begin{eqnarray}
     M^{ab}{}_{:ac}\epsilon^c{}_b  &=&
     \sin\theta\left(M^{\theta\phi}{}_{,\theta\theta}
                     + M^{\phi\phi}{}_{,\phi\theta}\right)
      - \csc\theta \left(M^{\theta\theta}{}_{,\theta\phi}
                     + M^{\phi\theta}{}_{,\phi\phi}\right)
       -\cot\theta\csc\theta M^{\theta\theta}{}_{,\phi}\nonumber\\
     &&\qquad + 5\cos\theta M^{\theta\phi}{}_{,\theta}
                    + 3\cos\theta M^{\phi\phi}{}_{,\phi}
                    + 3\left(\cos\theta\cot\theta -
                    \sin\theta\right) M^{\theta\phi},
\label{MderivsC}
\end{eqnarray}

\medskip
\noindent {\sl Exercise 4. Verify Eqs. (\ref{MderivsG}) and
(\ref{MderivsC}).}
\medskip

Any STF $2\times2$ tensor field can be written as the `gradient'
of some scalar field $A(\hat{n})$,
\begin{equation}
     A_{:ab}-\frac{1} {2} g_{ab}A^{:c}_c,
\end{equation}
plus the `curl' of some other scalar field $B(\hat{n})$,
\begin{equation}
     \frac{1}{2}(B_{:ac}\epsilon^c_b+B_{:bc}\epsilon^c_{a}).
\end{equation}
Just for comparison, a vector field is analogously decomposed as
\begin{equation}
     V_a=\nabla_aA+\epsilon_{ab}\nabla_bB.
\end{equation}
Since any scalar field on the sphere can be expanded in
spherical harmonics (e.g. for the temperature),
\begin{equation}
     {T(\hat{n}) \over T_0}=1+\sum_{l=1}^\infty\sum_{m=-l}^l
     a^{\rm T}_{(lm)}\,Y_{(lm)}(\hat{n}),
\label{Texpansion}
\end{equation}
where
\begin{equation}
      a^{\rm T}_{(lm)}={1\over T_0}\int d\hat{n}\,T(\hat{n})
       Y_{(lm)}^*(\hat{n}), 
\label{temperaturemoments}
\end{equation}
it follows that the polarization tensor can be expanded
in terms of basis functions that are gradients and curls of
spherical harmonics,
\begin{equation}
     {{\cal P}_{ab}(\hat{n})\over T_0} = \sum_{l=2}^\infty\sum_{m=-l}^l
     \left[ a_{(lm)}^{\rm G}
     Y_{(lm)ab}^{\rm G}(\hat{n}) + a_{(lm)}^{\rm C} Y_{(lm)ab}^{\rm C}
     (\hat{n}) \right].
\label{Pexpansion}
\end{equation}
The expansion coefficients are given by
\begin{equation}
     a_{(lm)}^{\rm G}={1\over T_0}\int \, d\hat{n} {\cal P}_{ab}(\hat{n})
                             Y_{(lm)}^{{\rm G} \,ab\, *}(\hat{n}), \qquad\qquad
     a_{(lm)}^{\rm C}={1\over T_0}\int d\hat{n}\, {\cal P}_{ab}(\hat{n})
                                      Y_{(lm)}^{{\rm C} \, ab\, *}(\hat{n}),
\label{defmoments}
\end{equation}
and
\begin{equation}
     Y_{(lm)ab}^{\rm G} = N_l
     \left( Y_{(lm):ab} - {1\over2} g_{ab} Y_{(lm):c}{}^c \right),
\label{Yplusdefn}
\end{equation}
\begin{equation}
     Y_{(lm)ab}^{\rm C} = { N_l \over 2}
     \left(\vphantom{1\over 2}
       Y_{(lm):ac} \epsilon^c{}_b +Y_{(lm):bc} \epsilon^c{}_a \right),
\label{Ytimesdefn}
\end{equation}
constitute a complete orthonormal set of basis functions for the
G and C components of the polarization.  The quantity,
\begin{equation}
     N_l \equiv \sqrt{ {2 (l-2)! \over (l+2)!}},
\label{Nleqn}
\end{equation}
is a normalization factor chosen so that
\begin{equation}
      \int d\hat{n}\,Y_{(lm)ab}^{{\rm G}\,*}(\hat{n})\,Y_{(l'm')}^{{\rm
      G}\,\,ab}(\hat{n})
      =\int d\hat{n}\,Y_{(lm)ab}^{{\rm C}\,*}(\hat{n})\,Y_{(l'm')}^{{\rm
      C}\,\,ab}(\hat{n})
      =\delta_{ll'} \delta_{mm'},\nonumber
\end{equation}
\begin{equation}
     \int d\hat{n}\,Y_{(lm)ab}^{{\rm G}\, *}(\hat{n})\,
     Y_{(l'm')}^{{\rm C}\,\,ab}(\hat{n})
     =0.
\label{norms}
\end{equation}
Also, we can integrate by parts to write alternatively,
\begin{equation}
     a_{(lm)}^{\rm G} = {N_l\over T_0}
     \int d\hat{n} \, Y_{(lm)}^*(\hat{n})\,
     {\cal P}_{ab}{}^{:ab}(\hat{n}),
\label{Gmomentseasy}
\end{equation}
\begin{equation}
     a_{(lm)}^{\rm C} = {N_l\over T_0}
     \int d\hat{n} \, Y_{(lm)}^*(\hat{n})\,
     {\cal P}_{ab}{}^{:ac}(\hat{n}) \epsilon_c{}^b.
\label{Cmomentseasy}
\end{equation}
Finally, since $\{{\rm T}, {\rm Q}, {\rm U}\}\in \Re$,
\begin{equation}
     a_{(lm)}^{\rm X\,*} =
     (-1)^m a_{(l,-m)}^{\rm X},
\label{almsymmetry}
\end{equation}
where ${\rm X}= \{{\rm T,G,C}\}$.

\medskip
\noindent {\sl Exercise 5.  Consider a vector field $V^a$ that
lives on the surface of a two-dimensional sphere.  Show that
this vector field can be written as
$$
     V_a = \sum_{l=1}^\infty \sum_{m=-l}^{l} \left[
     a_{(l m)}^{\rm G} Y_{(l m)a}^{\rm G} + a_{(l m)}^{\rm C}
     Y_{(l m)a}^{\rm C} \right],
$$
where $Y_{(l m)a}^{\rm G}$ and $Y_{(l m)a}^{\rm C}$ are
gradient and curl vector spherical harmonics.  Derive
expressions for these harmonics and show that they are
orthonormal.}
\medskip

The temperature/polarization power spectra are now
\begin{equation}
     \VEV{a_{(lm)}^{\rm X\,*}a_{(l'm')}^{\rm X'}} = C_l^{\rm X\rm
     X'}\delta_{ll'}\delta_{mm'},
\end{equation}
for $\rm X,\rm X'=\{T, G, C\}$. Of these six, two (TC and
GC) are zero if parity is conserved.\footnote{Our GG,CC, and TG
moments can be identified with the EE, BB, and TE moments of
Ref. \cite{zstensor} through $a_{(lm)}^{\rm G}= a_{lm}^{\rm E}/\sqrt{2}$
and $a_{(lm)}^{\rm C}= a_{lm}^{\rm B}/\sqrt{2}$.  Also, the
$C_l$ here reduce in the small-angle (large-$l$) limit with
those in Section 4 as long as the angles in the flat-sky limit
are given in radians; it is left as another exercise to the
reader to verify this statement.}

\begin{figure}[tbp]
\centerline{\psfig{file=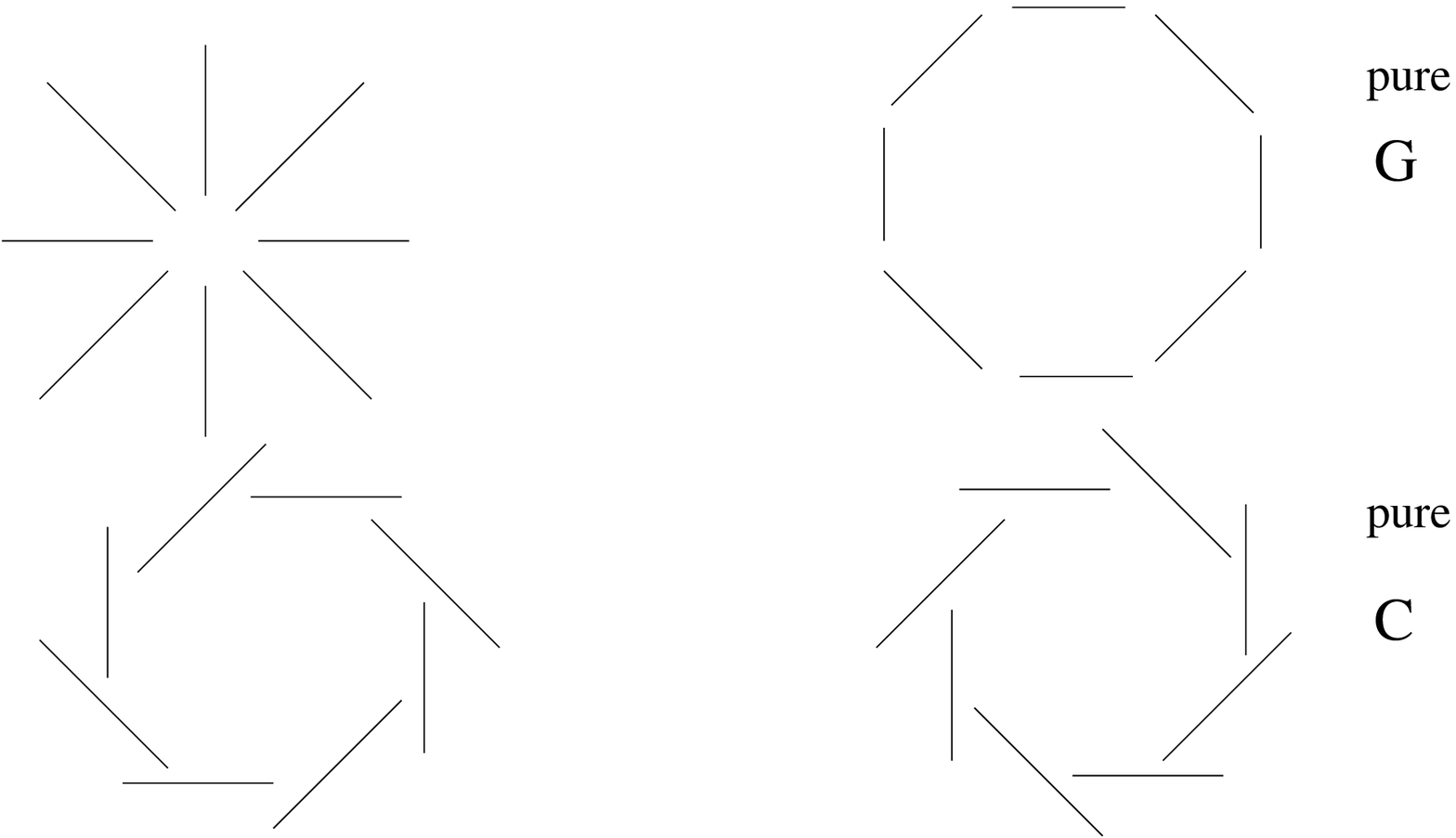,width=12cm}}
\bigskip
\fcaption{Examples of G (top) and C (bottom) polarization patterns.}
\label{fig:pureGC}
\end{figure}

For future reference, the $ Y_{(lm)ab}^{\rm G}$ and $ Y_{(lm)ab}^{\rm
C}$ are explicitly,
\begin{equation}
   Y_{(lm)ab}^{\rm G}(\hat{n})={N_l\over 2} \left( \begin{array}{cc}
   W_{(lm)}(\hat{n}) & X_{(lm)}(\hat{n}) \sin\theta\\
   \noalign{\vskip6pt}
   X_{(lm)}(\hat{n})\sin\theta & -W_{(lm)}(\hat{n})\sin^2\theta \\
   \end{array} \right),
\label{YGexplicit}
\end{equation}
and
\begin{equation}
   Y_{(lm)ab}^{\rm C}(\hat{n})={N_l\over 2} \left( \begin{array}{cc}
   -X_{(lm)}(\hat{n}) & W_{(lm)}(\hat{n}) \sin\theta \\
   \noalign{\vskip6pt}
   W_{(lm)}(\hat{n})\sin\theta & X_{(lm)}(\hat{n})\sin^2\theta \\
   \end{array} \right),
\label{YCexplicit}
\end{equation}
where
\begin{equation}
     W_{(lm)}(\hat{n}) = \left( {\partial^2 \over \partial\theta^2} -
     \cot\theta {\partial \over \partial\theta} +
     {m^2\over\sin^2\theta}\right)
     Y_{(lm)}(\hat{n}) = \left( 2{\partial^2 \over
     \partial\theta^2} + l(l+1) \right) Y_{(lm)}(\hat{n}),
\label{Wdefn}
\end{equation}
and
\begin{equation}
     X_{(lm)}(\hat{n}) = {2im \over \sin\theta}
     \left( {\partial \over \partial\theta} -
     \cot\theta \right) Y_{(lm)}(\hat{n}).
\label{Xdefn}
\end{equation}
In terms of the spin-2 harmonics ${}_{\pm2}Y_{(lm)}$
used in Refs. \cite{zstensor,huwhite},
\begin{equation}
     W_{(lm)}(\hat n) \pm i X_{(lm)}(\hat n) = \sqrt{{ (l+2)!
     \over (l-2)!}}\,_{\pm2}Y_{(lm)}.
\end{equation}
Note that if we replace $(Q,U)$ by $(U,-Q)$, then ${\rm G}\rightarrow
{\rm C}$ and ${\rm C}\rightarrow -{\rm G}$.  This tells us therefore, that a
pure-G polarization pattern becomes a pure-C pattern if we
rotate each polarization vector by $45^\circ$, and {\it vice
versa}.  Examples of G and C type polarization patterns are
shown in Fig. \ref{fig:pureGC}.

\section{Correlation functions on the sphere}

Correlation functions on the two-sphere are defined analogously
to those on the flat sky.  The line connecting the two points is
now replaced by the great arc connecting the two points \cite{kks}.  To
calculate the two-point correlation functions, it is
convenient to choose the north pole of the coordinate system to
be one of the points.  The temperature autocorrelation function
is then
\begin{eqnarray}
     C^{\rm TT}(\theta) &=& \VEV{ \frac{T(0,0)}{T_0}
     \frac{T(\theta,0)} {T_0} }\nonumber \\
     & =& \sum_{lml'm'} \VEV{a_{(lm)}^{{\rm T}\,*} a_{(l'm')}^{\rm
     T}} Y_{(lm)}^*(0,0)
       Y_{(l'm')}(\theta,0) \nonumber \\
     & =& \sum_{lml'm'} C_l^{\rm TT}  \delta_{ll'} \delta_{mm'}  \sqrt{ {2l+1
     \over 4 \pi} } \delta_{m0} Y_{(l'm')}(\theta,0) \nonumber \\
     & =& \sum_l\, {2l+1 \over 4 \pi}\, C_l^{\rm TT}\, P_l(\cos\theta).
\label{TTresult}
\end{eqnarray}
For polarization, we need a bit more algebra.  We begin by
deriving expressions for the functions that appear in our
definitions above of the tensor harmonics:
\begin{equation}
     W_{(lm)}(\hat{n}) = -2
     \sqrt{ {2l+1 \over 4\pi} {(l-m)! \over
     (l+m)!} } \, G_{(lm)}^+(\cos\theta)\, e^{im\phi},
\label{WGplus}
\end{equation}
\begin{equation}
     i X_{(lm)}(\hat{n}) =  -2
     \sqrt{ {2l+1 \over 4\pi} {(l-m)! \over
     (l+m)!} } \, G_{(lm)}^-(\cos\theta)\, e^{im\phi},
\label{XGminus}
\end{equation}
with
\begin{equation}
     G_{(lm)}^+(\cos\theta) \equiv - \left( {l-m^2 \over
     \sin^2\theta} + {1\over2} l(l-1) \right) P_l^m(\cos\theta)
     +(l+m) {\cos\theta \over \sin^2\theta}
     P_{l-1}^m(\cos\theta),
\label{Gplus}
\end{equation}
\begin{equation}
     G_{(lm)}^-(\cos\theta) \equiv {m\over\sin^2\theta}
     \Bigl( (l-1) \cos\theta
     P_l^m(\cos\theta) - (l+m)P_{l-1}^m(\cos\theta) \Bigr).
\label{Gminus}
\end{equation}
The Legendre functions have the following asymptotic limits,
\begin{equation}
     P_l^m(\cos\theta) \sim { (-1)^{(m+|m|)/2} \over 2^{|m|} |m|!}
     {(l+|m|)! \over (l-|m|)!} \, \theta^{|m|}, \qquad
     \theta\rightarrow0 \qquad\qquad (m\neq 0),
\end{equation}
\begin{equation}
     P_l^m(\cos\theta)\sim 1-\frac{1}{4}l(l+1)\theta^2, \qquad \theta
     \rightarrow0.
\end{equation}
We also note that at the north pole, $X_{lm}(\theta=0)\neq0$ and
$W_{lm}(\theta=0)\neq 0$ only for $m=2$:
\begin{equation}
     W_{(lm)}(0,0) = {1\over2} \sqrt{ {2l+1 \over 4\pi} { (l+2)! \over
     (l-2)!} } ( \delta_{m,2} + \delta_{m,-2}),
\label{Wzero}
\end{equation}
and
\begin{equation}
     X_{(lm)}(0,0) = {i\over2} \sqrt{ {2l+1 \over 4\pi} { (l+2)! \over
     (l-2)!} } ( \delta_{m2} - \delta_{m,-2}).
\label{Xzero}
\end{equation}

The QQ autocorrelation function is then defined as
\begin{equation}
     C^{\rm QQ}(\theta) = \VEV{\frac{Q_r(\hat{n}_1)} {T_0}
      \frac{Q_r(\hat{n}_2)}{T_0 }}_{\hat{n}_1\cdot
     \hat{n}_2=\cos\theta},
\label{Qcorrelation}
\end{equation}
where $Q_r$ is measured with respect to the great arc connecting
the two points. Let $Q(0,0)\equiv \lim_{\theta\rightarrow 0}
Q(\theta,0)$.  Then,
\begin{equation}
     C^{\rm QQ}(\theta) =  \VEV{ \frac{Q_r(0,0)}{T_0}
      \frac{Q_r(\theta,0)} {T_0}},
\label{Qcorrelation_1}
\end{equation}
or
\begin{eqnarray}
     C^{\rm QQ}(\theta) &=& \VEV{ \frac{Q(0,0)} {T_0}
       \frac{Q(\theta,0)} {T_0} } \nonumber \\
      &=& \sum_{lml'm'} N_l N_{l'}
     \Biggl\langle  \left[a_{(lm)}^{\rm G} W_{(lm)}(0,0) -
     a_{(lm)}^{\rm C} X_{(lm)}(0,0)\right]  \nonumber \\
      & & \times \left[a_{(l'm')}^{{\rm G}\,*}
     W_{(l'm')}^*(\theta,0) -
     a_{(l'm')}^{{\rm C}\,*} X_{(l'm')}^*(\theta,0) \right]
     \Biggr\rangle \nonumber \\
     & = & \sum_l \sqrt{2l+1 \over 8\pi}N_l [ C_l^{\rm G}(W_{(l2)}^* +
      W_{(l,-2)}^*) + iC_l^{\rm C}(X_{(l2)}^* - X_{(l,-2)}^*) ],\nonumber\\
     &=& \sum_{l}\frac{2l+1}{2\pi}N_l^2[C_l^{\rm G} G^{+}_{l2}
(\cos\theta)+C_l^{\rm C} G^{-}_{l2}(\cos\theta)],
\end{eqnarray}
where we have used in the first line,
\begin{eqnarray}
     Q(\hat{n})&=&2{\cal P}_{\theta\theta}(\hat{n})
      = T_0 \sum_{l=2}^\infty\sum_{m=-l}^l N_l
        \left[a_{(lm)}^{\rm G} W_{(lm)}(\hat{n})-a_{(lm)}^{\rm C}
        X_{(lm)}(\hat{n}) \right], \\
     U(\hat{n})&=&-2\csc\theta{\cal P}_{\theta\phi}(\hat{n})
     \nonumber \\
       &=&-T_0 \sum_{l=2}^\infty\sum_{m=-l}^l N_l
        \left[a_{(lm)}^{\rm G} X_{(lm)}(\hat{n})+a_{(lm)}^{\rm C}
        W_{(lm)}(\hat{n})\right].
\label{QUexpansion}
\end{eqnarray}
Similarly, the UU correlation function is
\begin{equation}
     C^{\rm UU}(\theta)=\sum_{l}\frac{2l+1}{2\pi}N_l^2
     \left[C_l^{\rm C} G^{+}_{l2}(\cos\theta)+C_l^{\rm G}
     G^{-}_{l2}(\cos\theta)\right],
\end{equation}
and the TQ correlation function is
\begin{equation}
     C^{\rm TQ}(\theta) =  \VEV{\frac{T(\hat{n}_1)}{ T_0}
     \frac{Q_r(\hat{n}_2)} 
     {T_0} }_{\hat{n}_1\cdot\hat{n}_2=\cos\theta}
        = \sum_l\, {2l+1 \over 4\pi}\, N_l \, C_l^{\rm TG}\,
        P_l^2(\cos\theta).
\label{TQresult}
\end{equation}
Note again that the TU and QU correlation functions vanish if parity is
conserved.

\medskip
\noindent {\sl Exercise 6. If parity is broken, do the TU and QU
correlation functions vanish?  If not, calculate them in terms
of the (parity-violating) power spectra.}
\medskip

\section{Calculation of Predicted Power Spectra}

Now that we have figured out how to describe the polarization
field properly, let's proceed to see what theoretical models
(e.g. primordial adiabatic density perturbations and
gravitational waves) predict about the polarization. Since these
calculations are extremely involved in practice, here we
only outline the calculation. We will, however, be able
to see precisely that gravitational waves produce a curl
component, while density perturbations do not.

To begin, we follow the pioneering paper of Polnarev
\cite{polnarev} to derive the angular distribution of photon
intensities in the presence of a gravitational wave (GW). To begin,
we suppose that the photons do not scatter. In this case, the
photon energies are affected only by the form of the metric.
Let us consider a single monochromatic plane-wave gravitational
wave, which appears as a tensor perturbation to the FRW metric,
\begin{equation}
     ds^2=a^2(\eta)d\eta^2-a^2(\eta)[dx^2(1+h_+)+dy^2(1-h_+)+dz^2],
\end{equation}
where $\eta$ is the conformal time and
\begin{equation}
     h_+(\vec{x},\eta)=h(\eta)e^{ik\eta}e^{-ikz},
\end{equation}
describes a plane wave propagating in the $\hat{z}$
direction. This is a linearly-polarized gravitational wave with
``+'' (rather than ``$\times$'') polarization.  Here $h(\eta)$ is the
amplitude; at early times when $k\eta\lesssim 1$,
$h(\eta)\simeq$const, but then $h(\eta)$ redshifts away when
$k\eta\gtrsim 1$ (see, e.g., Ref. \cite{Peeblesbook}).

\medskip
\noindent {\sl Exercise 7. Show that the amplitude $h_+(\eta)$ solves
$$
     \ddot h_+ + 2 {\dot a \over a} \dot h_+ + k^2 h_+ =0,
$$
where the dot denotes derivative with respect to conformal time
(actually, anisotropic stresses give something on the right-hand
side, but we will neglect them).  Then solve this equation for
the case of pure matter domination and for pure radiation
domination.  (Your solution should be consistent with a
gravitational-wave energy density $\rho_{GW} \propto a^{-4}$
with the scale factor $a(t)$ of the Universe when the mode
wavelengths are smaller than the horizon.)  Think about what the
spectrum of gravitational
waves should be in the Universe today, assuming an initially
scale-free spectrum.  You should find that the slope of the
spectrum changes at the scale that entered the horizon at
matter-radiation equality.  The spectrum is plotted in Fig. 3 of
Ref. \cite{caldwell}, and you should be able to understand the
result plotted there analytically.}
\medskip

With the metric
$\left[g_{\alpha\beta}=a^2\,{\rm diag}(1,-(1+h_+),-(1-h_+),-1)\right]$,
the zeroth component $P_0$ of the photon four-momentum is the
photon energy $E$ multiplied by the scale factor $P_0=aE$;
$P_0$ is constant if $h_+=0$. We take $h_+\ll 1$ for
small-amplitude gravitational waves.

Using the geodesic equation,
\begin{equation}
     \frac{d^2x^\mu}{d\lambda^2} = -\Gamma^\mu_{\alpha\beta}
     \frac{dx^\alpha}{d\lambda}\frac{dx^\beta}{d\lambda},
\end{equation}
\begin{equation}
     P^\alpha=\frac{dx^\alpha}{d\lambda},\:\:\:
     \frac{d}{d\lambda} = \frac{dx^0}{d\lambda} \frac{d}{dx^0} =
     \frac{d\eta}{d\lambda}\frac{d}{d\eta}=P^0\frac{d}{d\eta},
\end{equation}
and $g_{\mu\nu}P^\mu P^\nu=0$ for photons, we arrive at the
Sachs-Wolfe effect for this spacetime (to first order in $h$),
\begin{equation}
     \frac{1}{P_0}\frac{dP_0}{d\eta}=-\frac{1}{2}\frac{\partial
     h_+}{\partial \eta}(e^2_x-e^2_y),
\label{eqn:SW}
\end{equation}
where $e_x=(1-\mu^2)^{1/2}\cos\phi$ and
$e_y=(1-\mu^2)^{1/2}\sin\phi$ are the components of the
direction of the photon momentum, and $\mu=\cos\theta$ and
$\phi$ describe the photon direction.

\medskip
\noindent {\sl Exercise 8. Verify Eq. (\ref{eqn:SW}).}
\medskip

Replacing $P_0$ by the comoving photon frequency
$\nu=P_0=a\nu_{phys}=$const (in the unperturbed FRW spacetime),
we find that the gravitational wave
induces angular intensity variations of the form,
\begin{equation}
     \frac{d\nu}{\nu d\eta}=-\frac{1}{2}(1-\mu^2)\cos2\phi
     e^{-ikz}\frac{d}{d\eta}(he^{ik\eta}).
\end{equation}
Now let us consider Thomson scattering of these photons by
electrons.  Scattering will change this angular distribution and
it may induce polarization as well.  To include the effect of
Thomson scattering, we (following Polnarev 1986 \cite{polnarev})
consider an alternative set of ``Stokes parameters'', and
describe the state of the radiation propagating in any given
direction $\hat{n}$ by the ``vector'',
\begin{equation}
\tilde{I}=\left(
\begin{array}{c}
I_{\theta}\\
I_\phi\\
U\\
\end{array}
\right),
\end{equation}
where $I_\theta=a^2_\theta$, $I_\phi=a^2_\phi$,
$Q=I_\theta-I_\phi$, and $I=I_\theta+I_\phi$, and $Q$ and $U$ are measured
with respect to the $\hat{\theta}$-$\hat{\phi}$ axes on the plane
tangent to the sky at any given direction
$\hat{n}=(\theta,\phi)$.

The distribution function $\tilde{f}$ for photons is now also a
``vector''.  For example, in a homogeneous universe, it is
\begin{equation}
\tilde{f}=f_0(\nu)\left(
\begin{array}{c}
1\\
1\\
0\\
\end{array}
\right)=\tilde{f}_0(\theta,\phi),
\end{equation}
where
\begin{equation}
f_0(\nu)=\frac{1}{e^{h\nu/kT}-1}
\end{equation}
is the usual blackbody distribution function, and the form $(1 \:1\: 0)$
indicates no polarization.  In the presence of the gravitational
wave, there will be a perturbation, as determined above, so
\begin{equation}
\tilde{f}(\theta,\phi)=f_0\left[\left(
\begin{array}{c}
1\\
1\\
0\\
\end{array}
\right)
+\tilde{f}_1\right].
\end{equation}
Note that $\tilde{f}_1$ is a three-component ``vector'', and it is most
generally a function of conformal time $\eta$, position
$\vec{x}$, frequency $\nu$, and photon direction
$(\theta,\phi)=\hat{n}$.

As the Universe expands, photons get re-distributed in
frequency, polarization, and direction by the redshift due to
the gravitational wave and also by Thomson scattering.  Until
now we have considered only the gravitational redshift and
neglected scattering, which we now proceed to include. The time
evolution of the photon distribution function is determined by the
equation of radiative transfer, essentially the Boltzmann
equation for the photon distribution function,
\begin{equation}
     \frac{d\tilde{f}}{d\eta} = \frac{\partial\tilde{f}}
     {\partial\eta} + \hat{n}^i
     \frac{\partial\tilde{f}}{\partial x^i} +
     \frac{\partial\tilde{f}}{\partial\nu}
     \frac{\partial\nu}{\partial\eta} = g(\tilde{f}-\tilde{J}),
\label{eqn:boltzmann}
\end{equation}
where $g=\sigma_T n_e a$ is the scattering rate, $n_e$ the
electron density, and 
\begin{equation}
     \tilde{J} = \frac{1}{4\pi} \int_{-1}^{1} d\mu' \int_{0}^{2\pi} 
     d\phi' \tilde{\tilde{P}}(\mu,\phi,\mu',\phi')
     \tilde{f}(\eta,x^i,\nu,\mu',\phi'),
\end{equation}
where
\begin{equation}
\tilde{\tilde{P}}=\frac{3}{4}\left(
\begin{array}{ccc}
     \mu^2\mu'^2\cos2(\phi'-\phi) & -\mu^2\cos2(\phi'-\phi) &
     \mu^2\mu' \sin2(\phi'-\phi)\\
     -\mu'^2\cos2(\phi'-\phi)&\cos2(\phi'-\phi)&-\mu'\sin2(\phi'-\phi)\\
     -2\mu\mu'^2\sin2 (\phi'-\phi) & 2\mu \sin2(\phi'-\phi) & 2
     \mu\mu'\cos2(\phi'-\phi)
\end{array}
\right)
\end{equation}
is the scattering matrix.
Eq. (\ref{eqn:boltzmann}) says that the total time derivative of
the photon distribution function (written as an explicit time
derivative plus the time evolution due to photon motion) is
given by gravitational redshift (the first term on the
right-hand side) in the perturbed spacetime plus the change due
to Thomson scattering (the second term on the right-hand side)
of photons from electrons.  The quantity $\tilde{J}$ is
the angular intensity-polarization distribution that arises
after the radiation has been Thomson scattered once from an initial
distribution $\tilde{f}$.  The scattering matrix $\tilde{\tilde
P}$ looks messy, but it simply tells us how an initial
intensity-polarization pattern is re-arranged after Thomson
scattering once from unpolarized electrons.  It is
straightforward to derive it from the angular-polarization
dependence of Thomson scattering,
\begin{equation}
     \frac{d\sigma_T}{d\Omega}\propto|\hat{\epsilon}\cdot\hat{\epsilon}'|^2;
\end{equation}
see Refs. \cite{polnarev} and \cite{Chandrasekhar} for details.

The gravitational wave imprints at first an angular
intensity pattern,
\begin{equation}
     \frac{1}{\nu}\frac{d\nu}{d\eta}=-\frac{1}{2}(1-\mu^2)\cos2\phi
     e^{-ikz}\frac{d}{d\eta}(he^{ik\eta}).
\end{equation}
This intensity pattern is then altered when the photons undergo
Thomson scattering.  Before Thomson scattering, the
intensity-polarization distribution function is the unperturbed
distribution $\tilde f_0$ plus a small perturbation proportional to
\begin{equation}
\tilde a \equiv \frac{1}{2} \left(
\begin{array}{c}
1\\
1\\
0
\end{array}
\right)(1-\mu^2)\cos2\phi,
\end{equation}
due to the gravitational wave.  After Thomson scattering, the
angular intensity-polarization distribution is altered; the
scattering matrix above then introduces a second component to
the perturbation proportional to
\begin{equation}
\tilde b \equiv \frac{1}{2}\left(
\begin{array}{c}
(1+\mu^2)\cos2\phi\\
-(1+\mu^2)\cos2\phi\\
4\mu\sin2\phi
\end{array}
\right),
\end{equation}
One might guess that subsequent scatterings would
add further more complicated angular dependence to the
distribution function.  However, as first noted by Polnarev
\cite{polnarev}, Thomson scattering of a distribution function
of the form $\tilde b$ returns a distribution function of the
form $\tilde a$.  In other words, the basis functions $\tilde a$
and $\tilde b$ form a closed basis under Thomson scattering.  
This is simply a consequence of the time-reversal
invariance of the scattering process:  If Thomson scattering
converts a distribution $\tilde a$ to a distribution $\tilde b$,
then it should turn a distribution $\tilde b$ into a
distribution $\tilde a$.

The linearized solution to the Boltzmann equation, including
fully the effects of gravitational redshift and Thomson
scattering, in the presence of a gravitational wave must
therefore be of the form
\begin{equation}
     \tilde{f}=\tilde{f}_0+e^{-ikz+ik\eta}\tilde{f}_1,
\end{equation}
where the perturbation is
\begin{equation}
     \tilde{f}_1 = \alpha(\eta,\nu_0,\mu) \tilde{a}+
     \beta(\eta,\nu_0,\mu)\tilde{b}.
\end{equation}
Here, $\alpha(\eta,\nu_0,\mu)$ and $\beta(\eta,\nu_0,\mu)$ are
coefficients that must be determined by solution of the
Boltzmann equation.

Defining an ``anisotropy'' $A=\xi(1-\mu^2)$ with
$\xi\equiv\alpha+\beta$, and ``polarization''
$\Pi=\beta(1+\mu^2)$ (which is nonzero only if there is
polarization), the radiative-transfer equation can be
re-written,
\begin{eqnarray}
     \dot{\xi}+[ik(1-\mu)+g]\xi & =
     &\frac{\nu_0}{f_0} \frac{df_0(\nu_0)}{d\nu_0}
     \left(\dot{h}+ikh\right),\nonumber\\
     \dot{\beta}+[ik(1-\mu)+g]\beta & = &\frac{3}{16}g\int
     d\mu'[(1+\mu'^2)^2\beta-\frac{1}{2} \xi (1-\mu'^2)^2].
\label{eqn:rewritten}
\end{eqnarray}
The first equation generates a temperature fluctuation from the
gravitational wave, and the second generates polarization
through scattering of that temperature
fluctuation.\footnote{Note that the factor $ik(1-\mu)$ that
appears in Eqs. (\ref{eqn:rewritten}) often appears elsewhere
simply as $i k \mu$.  The reason traces back to the fact that
Polnarev writes $h_+ \propto h(\eta)e^{ik\eta}$ and $\tilde f
=\tilde f_0 + \tilde f_1 e^{-ikz+ik\eta}$ while other authors
usually write$h_+\propto h(\eta)$ and $\tilde f = \tilde f_0 +
\tilde f_1$.  We thank J. Pritchard for clarifying this point.}

Eqs. (\ref{eqn:rewritten}) are a set of coupled partial
differential equations for the time ($\eta$) and angular ($\mu$)
dependence of the distribution functions $\xi(\eta,\mu)$ and
$\beta(\eta,\mu)$.  In practice, these are solved numerically by
Legendre transforming  $\xi(\eta,\mu)\rightarrow \xi_l(\eta)$
and $\beta(\eta,\mu)\rightarrow \beta_l(\eta)$ through
\begin{equation}
     \xi_l(\eta) = \frac{1}{2}\int_{-1}^1 \,d\mu\, \xi(\eta,\mu)P_l(\mu) ,
\end{equation}
\begin{equation}
     \beta_l(\eta) =\frac{1}{2}\int_{-1}^1\, d\mu\, \beta(\eta,\mu)P_l(\mu) .
\end{equation}
The partial differential equations for $\xi(\eta,\mu)$ and
$\beta(\eta,\mu)$ then become an infinite set\footnote{These are
truncated at some large value of $l$; the truncation procedure
is not trivial; see, e.g., Ref.~\cite{mabert}.} of coupled differential equations for
$\xi_l(\eta)$ and $\beta_l(\eta)$ which are propagated
numerically from some very early time (where the solutions are
analytically determined in the so-called
tight-coupling approximation) to the present time.

\medskip
\noindent{\sl Exercise 9.  Derive the differential equations for
$\xi_l(\eta)$ and $\beta_l(\eta)$; this is not necessarily an
easy problem.  If you want to go further, you can determine the
early-time solutions to these equations using the tight-coupling
approximation, in which the scattering rate $g=n_e \sigma_T a$
is assumed to be huge.}
\medskip

We will not discuss the numerical techniques (which are highly
nontrivial) here, but will show results  a bit later. For our
purposes, we simply need to know that the angular polarization
pattern induced by this gravitational wave can now be written,
\begin{eqnarray}
     Q(\theta,\phi) & = & \frac{1}{4}T_0 \sum_l(2l+1)
     P_l(\cos\theta)(1+\cos^2\theta)\cos2\phi\:\xi_l,\nonumber\\ 
     U(\theta,\phi) & = & \frac{1}{4} T_0\sum_l(2l+1)
     P_l(\cos\theta)2\cos\theta\sin2\phi\:\xi_l.
\label{eqn:QUintermsofl}
\end{eqnarray}
We then get a polarization tensor,
\begin{eqnarray}
     P^{ab}(\theta,\phi) = & &\frac{T_0}{8}
     \sum_l(2l+1)P_l(\cos\theta)\xi_l\nonumber \\
     & & \times\left(
     \begin{array}{cc}
     (1+\cos^2\theta)\cos2\phi & -2\cot\theta \sin2\phi \\
     -2\cot\theta \sin2\phi    & -(1+\cos^2\theta)\csc^2\theta\cos2\phi
\end{array}
\right).
\label{eqn:Pab}
\end{eqnarray}

\medskip
\noindent {\sl Exercise 10. Verify
Eq. (\ref{eqn:QUintermsofl}).  This should be easy.}
\medskip

If we now expand Eq. (\ref{eqn:Pab}) in tensor spherical
harmonics, the resulting tensor-harmonic coefficients are
\begin{equation}
     a^{\rm G}_{lm}=\frac{1}{8}N_l\sum_j(2j+1)\xi_j\int
     d\hat{n}Y^*_{lm}(\hat{n})M^{ab}_{(j):ab}(\hat{n}),
\end{equation}
which after a little bit of algebra becomes
\begin{eqnarray}
     a^{\rm G}_{lm} = & & \frac{1}{8}  (\delta_{m2}+\delta_{m,-2})
     \sqrt{2\pi(2l+1)} \nonumber \\
     \times & & \left[\frac{(l+2)(l+1)\xi_{l-2}}{(2l-1)(2l+1)}+\frac{6
     l(l+1) \xi_{l}}{(2l+3)(2l-1)} + \frac{l(l-1)\xi_{l+2}}
     {(2l+3)(2l+1)}\right].
\end{eqnarray}
Likewise,
\begin{eqnarray}
     a^{\rm C}_{lm}&=&\frac{1}{8}N_l\sum_j(2j+1)\xi_j\int
     d\hat{n}Y^*_{lm}(\hat{n})M^{ab}_{(j):ac}(\hat{n})\epsilon^c_b\\
     & = & \frac{-i}{4} \sqrt{\frac{2\pi}{(2l+1)}}
     (\delta_{m2}-\delta_{m,-2})[(l+2)\xi_{l-1}+(l-1)\xi_{l+1}].
\end{eqnarray}
We have thus shown explicitly that both the G and C components are
nonzero for a gravitational wave.

We get the contributions to the power spectra $C^{\rm GG}_l$ and
$C^{\rm CC}_l$ from this particular gravitational wave (in the
$\hat{z}$ direction with `+' polarization) from
\begin{eqnarray}
     C_l^{\rm GG} &=& \frac{1}{2l+1} \sum_m|a^G_{lm}|^2 \nonumber
     \\
     &=& \frac{\pi}{16}
     \left[\frac{(l+2)(l+1)\xi_{l-2}}{(2l-1)(2l+1)}+\frac{6
     l(l+1)\xi_{l}} {(2l+3)(2l-1)}+\frac{l(l-1) \xi_{l+2}}
     {(2l+3)(2l+1)}\right]^2
\end{eqnarray}
and similarly for $C_l^{\rm CC}$.  Summing over all Fourier modes,
$\int d^3k/(2\pi)^3$, and over both polarization states, the
final result for $C_l^{\rm GG}$ is
\begin{equation}
     C_l^{\rm GG}=\frac{1}{16\pi}\int k^2\, dk
     \left[ \frac{(l+2)(l+1)\xi_{l-2}}{(2l-1)(2l+1)}+\frac{6
     l(l+1)\xi_{l}} {(2l+3)(2l-1)} +
     \frac{l(l-1)\xi_{l+2}}{(2l+3)(2l+1)}\right]^2,
\end{equation}
and similarly for $C_l^{\rm CC}$.
Note, finally, that the cross-correlation power spectrum vanishes
\begin{equation}
     C_l^{\rm GC}= \sum_{m=-l}^{m=l}\frac{a^{{\rm
     G}*}_{lm}a^{\rm C}_{lm}}{2l+1}=0,
\end{equation}
as it should.  This is because $a_{(lm)}^{\rm G} \propto
(\delta_{m,2}+\delta_{m,-2})$, while $a_{(lm)}^{\rm C} \propto
(\delta_{m,2}-\delta_{m,-2})$.

\begin{figure}[tbp]
\centerline{\psfig{file=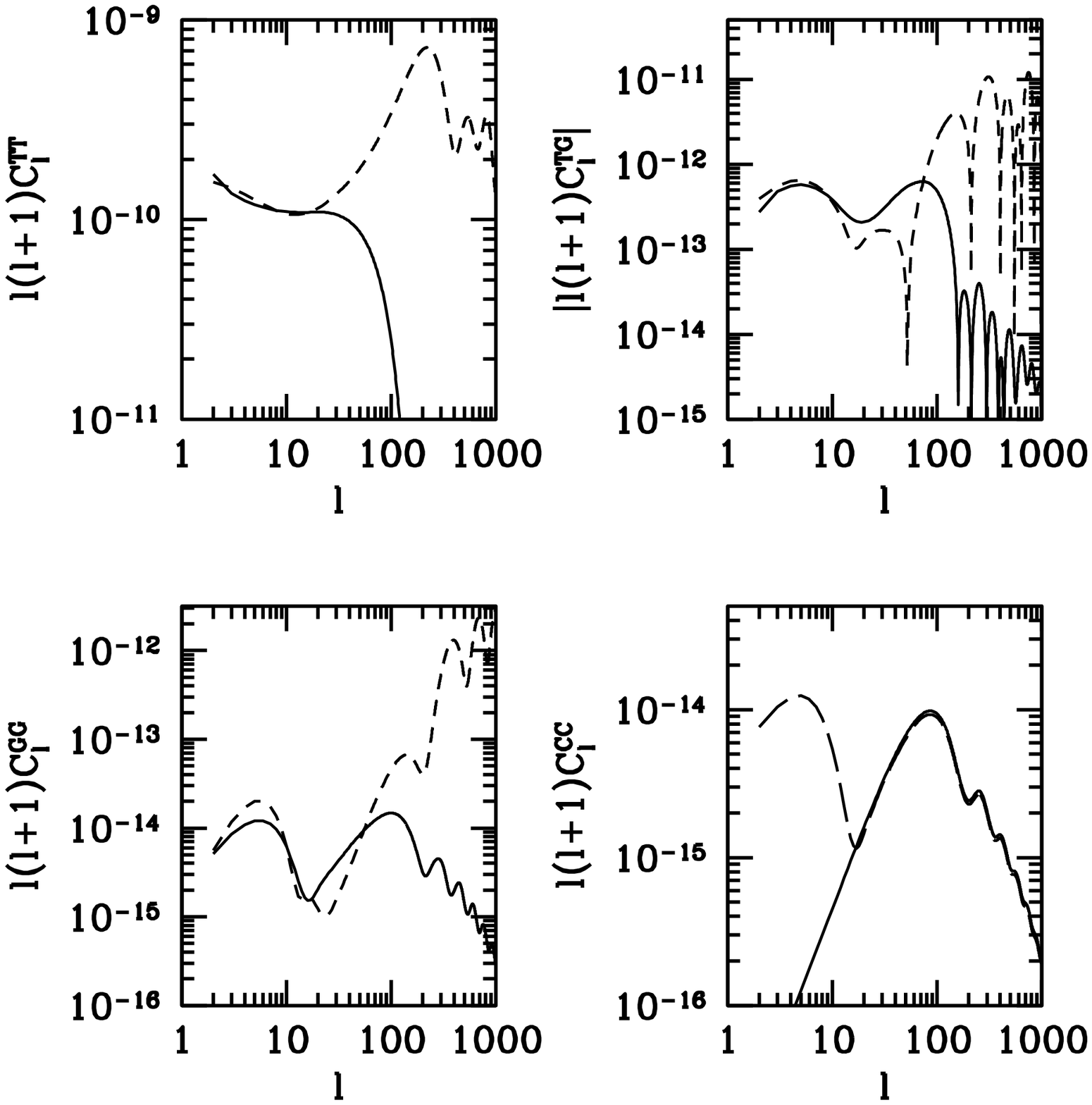,width=13cm}}
\bigskip
\fcaption{Power spectra of temperature and polarization (G and C) and
     cross-spectrum TG for scalar perturbations (dotted line)
     and perturbations due to gravitational waves (solid
     line). Note that the power spectrum of `curl' component C,
     due to scalar perturbations, is missing, and the dashed line
     corresponds a reionized model with optical depth $\tau=0.166$
     to the surface of last scattering.}
\label{fig:ps}
\end{figure}

Now what about scalar (density) perturbations?  By following
steps similar to those above, one can show that they do not
produce a curl component in the CMB polarization.  Here we only
sketch the calculation and leave out details.
Again, consider a single Fourier mode of the density field in
the $\hat{z}$ direction.  Then the Sachs-Wolfe effect induces an
intensity variation proportional to
$(\cos^2\theta-1/3)$, so now
\begin{equation}
     \tilde{f}_1 = \alpha(\eta,\nu_0,\mu) \tilde{a}
     +\beta(\eta,\nu_0,\mu)\tilde{b},
\end{equation}
with
\begin{equation}
\tilde{a}=\left(\mu^2-\frac{1}{3}\right)\left(
\begin{array}{c}
1\\
1\\
0
\end{array}
\right),\:\:\:\:\:
\tilde{b}=(1-\mu^2)\left(
\begin{array}{c}
1\\
-1\\
0
\end{array}
\right).
\end{equation}
We thus find that for $\vec{k}||\hat{z}$, $U(\hat{n})=0$, so
\begin{equation}
M_{(j)}^{ab}(\theta,\phi)=\sin^2\theta P_j(\cos\theta)\left(
\begin{array}{cc}
1 & 0\\
0 & \frac{-1}{\sin^2\theta}
\end{array}
\right).
\end{equation}
One finds $a^{\rm G}_{lm}\neq 0,$ but $a^{\rm C}_{lm}= 0$.  This
follows because $M^{ab}_{:ac}\epsilon^c_b=0$ which follows since
$M^{ab}_{:ac}$ is diagonal and independent of $\phi$.
Therefore, if one detects a curl component in the CMB, it is a
signature for primordial gravitational waves (see
Fig. \ref{fig:ps}).

\medskip
\noindent {\sl Exercise 11.  Show that if the gravitational-wave background is
composed entirely of gravitational waves with right-handed
circular polarization (and no left-handed gravitational waves)
that $C_l^{\rm TC}\neq0$; i.e., that there is a
cross-correlation between the CMB temperature and the curl
component of the polarization.  (Hint: Consider a single
circularly-polarized gravitational wave propagating in the ${\bf 
\hat z}$ direction; from Ref. \cite{lue}.)}
\medskip

\section{Comments on power spectra}

The power spectra shown in Fig. \ref{fig:ps} show a variety of
features.  The origin of these features, as well as their
dependence on cosmological parameters, has been the subject of
much study.  We will not discuss them in detail but refer the
reader to the reviews and books listed in the Introduction for
more details.  Here, we point out only a few interesting and
relevant features.

\begin{itemize}

\item There are acoustic peaks in the GG power spectrum for density
perturbations. These are out of phase from the peaks in the TT power
spectrum. The temperature fluctuation is due primarily to
density fluctuations at the
surface of last scattering, and secondarily to the peculiar
velocity at the surface of last scattering. The peaks in $C^{\rm
TT}_l$ are due to the density perturbations; the troughs are
filled in by peculiar velocities, which oscillate (just as in an
ordinary harmonic oscillator) out of phase by $90^\circ$.

Before recombination, the electrons and photons are tightly coupled,
so the electrons see no quadrupole in the photon intensity and thus produce
no polarization. Just before recombination, the photons begin to have longer
path lengths as they begin to decouple. Since this nonequilibrium process
depends on the time derivative of the baryon density (just like
the velocity does), the polarization is out of phase with the
density, and in phase with the velocity \cite{zd}.

\item The power spectra due to gravitational waves drop
precipitously for $l \gtrsim 100$. This is because on smaller
scales, the gravitational waves have entered the horizon  and
had time for the their amplitudes to redshift away by the time
of recombination.

\item Reionization induces a large bump in $C_{l}^{\rm GG}$,
$C_{l}^{\rm CC}$, and $C_l^{\rm TG}$ at $l\lesssim 10$
\cite{zaldarriaga}. This is simply due to scattering of the
quadrupole by reionized gas.  A rough estimate for the $l$ of
this peak can be obtained by assuming an Einstein--de-Sitter
Universe and noting then that
\begin{equation}
l_{\rm reion}\sim
     200\left(\frac{z_{\rm reion}}{z_{\rm rec}}\right)^{1/2}
     \frac{1}{\sqrt{3}}\sim10\:\:\:   {\mathrm for}\:\:\:
     z_{\rm reion}\sim10,
\end{equation}
where $l\sim200$ is the acoustic-peak location, and we have used
the approximation that the sound speed $c_s\simeq c/\sqrt{3}$ in
the primordial plasma is dominated by radiation.

\item $C^{\rm GG}_l$ from density perturbations peaks at $l\sim
1000$, as opposed to $l\sim 200$ for $C^{\rm TT}_l$: there is
more power on small scales in the polarization.  This is because
polarization induced by a particular Fourier
mode of the primordial density field depends on the gradient of
that density field.

\item The amplitude of $C^{\rm CC}_l$ is proportional to the
square of the amplitude of the gravitational-wave background,
which is fixed by the height of the inflaton potential during
inflation.  This can be quantified by 
\begin{equation}
     J\equiv 6C_2^{\rm TT,tensor}\simeq 10^{-10}
     \frac{V}{(3\times10^{16}\, {\mathrm GeV})^4},
\end{equation}
where $V$ is the inflaton-potential height during inflation, and
$C_2^{\rm TT,tensor}$ is the tensor (gravitational-wave)
contribution to the temperature quadrupole.  Since the latter is
$\sim 10^{-10}$, we already known that the energy scale of
inflation is $V^{1/4}\lesssim 3\times10^{16}\,{\mathrm GeV}$. The
curl component of the CMB polarization induced by gravitational
waves is thus proportional to the scale-height of inflation:
\begin{equation}
C^{\rm CC}_l\propto V.
\end{equation}
Therefore, detection of a curl component in the CMB due to
gravitational waves would provide not only a ``smoking gun'' for
inflation, but it would also tell us the height of the inflaton
potential during inflation, and thus provide some hints as to
the new physics responsible for inflation.  In practice, any
realistic CMB polarization experiment will be able to
detect the GW-induced CMB curl component only if
$V\gtrsim10^{15}$ GeV; in other words, only if inflation took
place at the energy scale of grand unification.

\item $C^{\rm CC}_l$ from gravitational waves peaks at
$l\sim100$ or at $\sim 2^\circ$. If there is no other sources of
a curl component (see below), then detection of the curl
component is not cosmic-variance limited.

\item Gravitational waves contribute a roughly scale-invariant
temperature power spectrum $C_l^{\rm TT}$ at low $l$, just like
density perturbations. However, cosmic variance removes our ability to detect
an excess of $\lesssim 10\%$ over the density-perturbation
contribution to $C_l^{\rm TT}$, even if we could predict
theoretically the density-perturbation amplitude.

\item If there is no other source of a curl component, and if
the large-angle polarization bump due to reionization is not
big, then the optimal survey strategy \cite{jkw} entails a deep
integration on a $\sim 5^\circ\times  5^\circ$ patch of sky,
with an angular resolution $\lesssim 1^\circ$ fwhm.  This is the
strategy of experiments like BICEP and QUIET.  We now know from
the WMAP detection of a large-angle temperature-polarization
correlation that the reionization bump is probably quite big.
In this case, an all-sky experiment with roughly the same instrumental
sensitivity (noise-equivalent temperature) might have similar
sensitivity to inflationary gravitational waves.

\end{itemize}

\medskip
\noindent {\sl Exercise 12. Scale-invariant spectra of density
perturbations and
of gravitational waves each produce a nearly scale-invariant
spectrum, $l(l+1) C_l^{\rm TT,tensor} \simeq$constant, in the CMB at
large scales (i.e., $l\lesssim20$).  Suppose that most of the
observed anisotropy at large scales comes from density
perturbations, and pretend that we knew precisely the amplitude of the
density-perturbation power spectrum from modeling smaller-scale
fluctuations, other observations, or divine inspiration.  If so,
there would be a limit, set by cosmic variance, to the smallest
gravitational-wave amplitude (quantified by, e.g., $6 C_2^{\rm
TT,tensor}$, the gravitational-wave contribution to the
temperature quadrupole moment) that could be determined from the
measured CMB power
spectrum.  Estimate the smallest detectable value of $C_2^{\rm
TT,tensor}$.  You will need to start by showing that the $1\sigma$
cosmic-variance error with which each $C_l^{\rm TT}$ can be
measured is $\sqrt{2/(2l+1)}C_l^{\rm TT}$.}
\medskip

\section{Gravitational Lensing (Cosmic Shear) and CMB Polarization}

Above we showed that density perturbations do not induce a curl
in the polarization, and thus concluded that detection of a curl
in the CMB polarization automatically implies detection of
gravitational waves. However, that derivation assumed only
linear perturbations, in which each Fourier mode of the density
field is considered independently. If more than one Fourier mode
of the density field is considered, then different modes can
interact and produce a curl, even without gravitational
waves. Since $\delta\sim10^{-5}$ at the last-scattering surface, this
density-perturbation--induced curl component should be small.

We do know, however, that the observed CMB
temperature-polarization map will be distorted by \emph{cosmic
shear} (CS), gravitational lensing by large-scale mass
inhomogeneities between us and the surface of last scattering.  The
effect of cosmic shear is to displace the temperature and
polarization from a given direction $\vec \theta$ at the surface
of last scattering to an adjacent position, $\vec\theta +
\delta\vec\theta$:
\begin{equation}
\left(\begin{array}{c}
T\\
Q\\
U\\
\end{array}
\right)_{obs.}(\vec{\theta})
=\left(\begin{array}{c}
T\\
Q\\
U\\
\end{array}
\right)_{ls}(\vec{\theta}+\delta\vec{\theta}) 
\simeq\left(\begin{array}{c}
T\\
Q\\
U\\
\end{array}
\right)_{ls}(\vec{\theta})+\delta\vec{\theta}\cdot\nabla\left(\begin{array}{c}
T\\
Q\\
U\\
\end{array}
\right)_{ls}(\vec{\theta}),
\end{equation}
where $\delta\vec{\theta} =\nabla\varphi$ is the cosmic-shear
displacement angle, and
\begin{equation}
     \varphi(\hat{n})=-2\int_0^{r_{ls}}dr
     \frac{d_{A}(r_{ls},r)}{d_A(r,0)d_A(r_{ls},0)}\Phi(r,\hat{n}r)
\end{equation}
is the projection of the gravitational potential
$\Phi(r,\hat{n}r)$ (obtained from the mass distribution through
the Poisson equation) along the line of sight, and $d_A(r_1,r_2)$ is
the angular-diameter distance corresponding to the comoving
radial coordinate $r_1$ by an observer at $r_2$.

Noting that
\begin{equation}
     \frac{l_x^2-l_y^2}{l_x^2+l_y^2}=
     \cos2\phi_{\vec l},\qquad
     \frac{2l_xl_y}{l_x^2+l_y^2}=\sin2\phi_{\vec l},
\end{equation}
our previous transformation [cf.,
Eq. (\ref{eqn:GCFouriercomponents})] between $(Q,U)$ and $(G,C)$
is (using $G$ and $C$ as a shorthand for ${\tilde P}_{\rm G}$
and ${\tilde P}_{\rm C}$, respectively),
\begin{equation}
\left(
\begin{array}{c}
G\\
C\\
\end{array}
\right)(\vec{l})
=\frac{1}{2}
\left(
\begin{array}{cc}
\cos2\phi_{\vec l} & \sin2\phi_{\vec l}\\
\sin2\phi_{\vec l} & -\cos2\phi_{\vec l} \\
\end{array}
\right)
\left(
\begin{array}{c}
Q\\
U\\
\end{array}
\right)(\vec{l}).
\end{equation}
Let us suppose the polarization field at the surface of last scattering
has no curl; then this relation can be inverted to give
\begin{equation}
     Q(\vec{l})=2G(\vec{l})\cos2\phi_{\vec l},\:\:\:\:\:
     U(\vec{l})=-2G(\vec{l})\sin2\phi_{\vec l},
\end{equation}
and
\begin{equation}
     \nabla Q(\vec\theta) = -2i\, \int\,
     \frac{d^2\vec{l}}{(2\pi)^2}\, 
     G(\vec{l})\, \cos2\phi_{\vec l}\, 
     \vec{l}\, e^{-i\vec{l}\cdot\vec{\theta}},
\end{equation}
and similarly for $\nabla U(\vec\theta)$ with $\cos\rightarrow
-\sin$.  The displacement angle is likewise
\begin{equation}
     \delta \vec\theta(\vec\theta)=-i\int\, \frac{d^2\vec{l}}
     {(2\pi)^2}\, \varphi(\vec{l})\,
     e^{-i\vec{l}\cdot\vec{\theta}}\, \vec{l}.
\end{equation}
Thus, the perturbation to $Q$ and $U$ induced by gravitational waves is
\begin{equation}
     \delta Q(\vec{\theta})=\nabla Q\cdot\nabla
     \varphi=\int\frac{d^2\vec{l}}{(2\pi)^2}
     e^{-i\vec{l}\cdot\vec{\theta}}(\nabla Q\cdot\nabla
     \varphi)_{\vec{l}},
\end{equation}
where
\begin{equation}
     \delta Q(\vec{l})\equiv (\nabla Q\cdot\nabla
     \varphi)_{\vec{l}} = 2 \int \frac{d^2\vec{l}_1}{(2\pi)^2}
     [\vec{l}_1\cdot(\vec{l} - \vec{l}_1)] G(\vec{l}_1)
     \varphi(\vec{l}-\vec{l}_1)\cos2\phi_{\vec{l}_1},
\end{equation}
\begin{equation}
     \delta U(\vec{l})\equiv (\nabla U\cdot\nabla
     \varphi)_{\vec{l}}=-2\int\frac{d^2\vec{l}_1}
     {(2\pi)^2}[\vec{l}_1\cdot(\vec{l}-\vec{l}_1)]G(\vec{l}_1)
     \varphi(\vec{l}-\vec{l}_1)\sin2\phi_{\vec{l}_1}.
\end{equation}
Although the original map had (by assumption) no curl, the lensed map does:
\begin{eqnarray}
     C(\vec{l})&=&\frac{1}{2}[\sin2\phi_{\vec{l}\,}Q(\vec{l})-
     \cos2\phi_{\vec{l}\,}U(\vec{l})]\nonumber \\
     &=&\int\frac{d^2\vec{l_1}}{(2\pi)^2}
     [\vec{l}_1\cdot(\vec{l}-\vec{l}_1)]G(\vec{l}_1)\varphi(\vec{l}-\vec{l}_1)
     \sin2(\phi_{\vec{l}} -\phi_{\vec{l}_1})\nonumber\\
     &=&\int\frac{d^2\vec{l_1}}{(2\pi)^2}
     [\vec{l}_1\cdot(\vec{l}-\vec{l}_1)]G(\vec{l}_1)
     \varphi(\vec{l}-\vec{l}_1)\sin2\phi_{\vec{l}_1}.
\end{eqnarray}
Thus, our earlier claim that detection of a curl component
constitutes detection of a gravitational-wave background is not
entirely valid, as we have just shown explicitly that a curl
component can be induced by cosmic shear \cite{zs}.

Using the power spectrum
$\VEV{|\varphi_{\vec{l}}|^2} = C_l^{\varphi\varphi}$ for the
projected potential, the power spectrum for the curl induced by
lensing is,
\begin{equation}
     C_l^{\rm CC}=\int\frac{d^2\vec{l_1}}{(2\pi)^2}
     [\vec{l}_1\cdot(\vec{l}-\vec{l}_1)]^2\sin^22
     \phi_{\vec{l}_1} C^{\varphi\varphi}_{|\vec{l}-\vec{l}_1|}
     C_{{l_1}}^{\rm GG}.
\end{equation}

\begin{figure}[tbp]
\centerline{\psfig{file=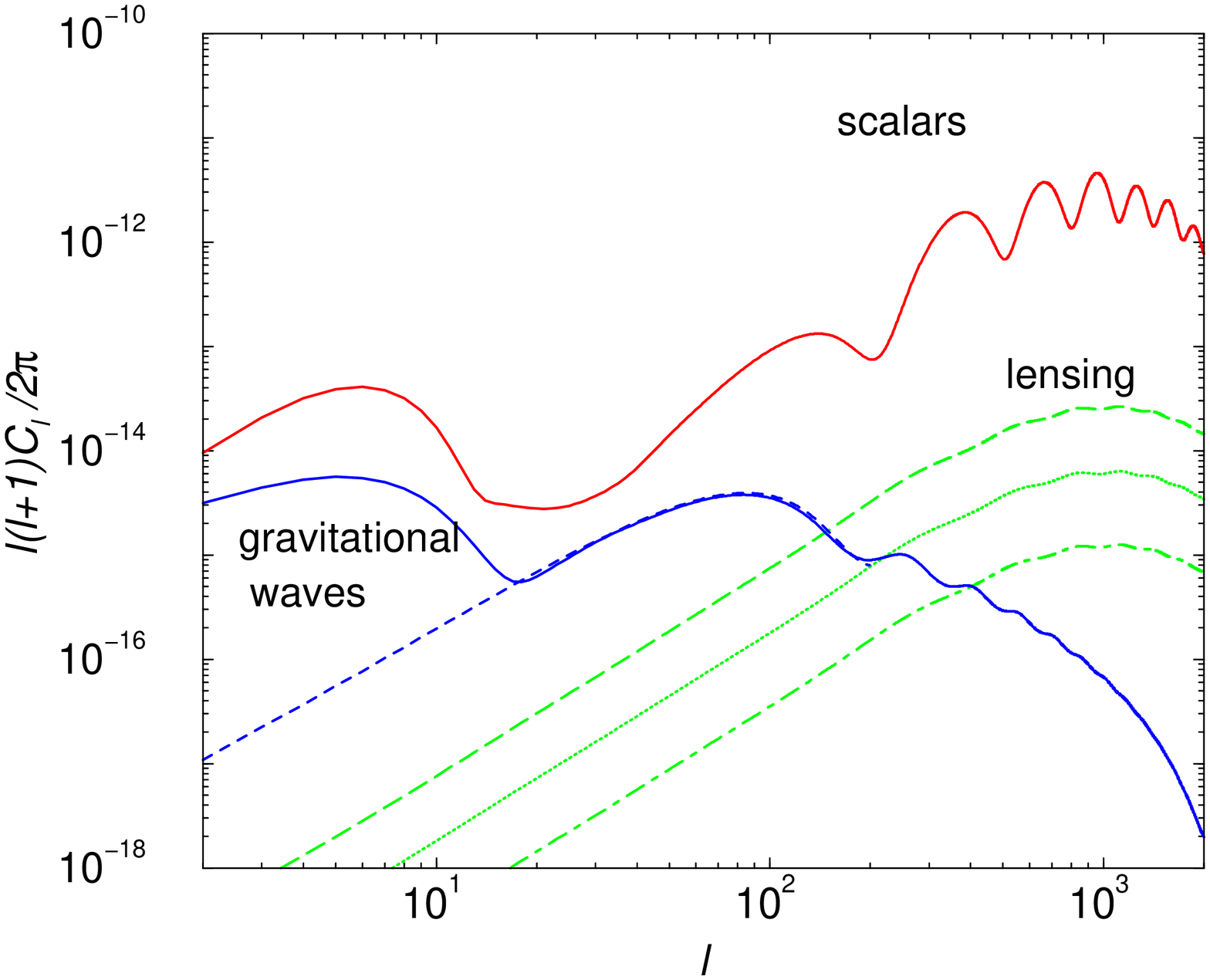,width=12cm,angle=-90}}
\bigskip
\fcaption{Contributions to the CMB polarization power spectra.
     The upper (blue) solid ``gravitational waves'' curve is the
     curl component
     due to gravitational waves in the presence of reionization
     with an optical depth $\tau=0.17$ (note the large-angle
     bump) and the associated (blue) dashed curve is that with no  
     reionization.  The amplitude of this curve is for the
     largest inflaton-potential height
     ($V\simeq3.5\times10^{16}$ GeV) allowed by COBE; note that
     $C_l^{\rm CC,GW}\propto V$, so the amplitude of this curve
     will be reduced accordingly if $V$ is reduced.  The
     short-dash (green) ``lensing'' curve is
     the curl power spectrum induced by cosmic shear (weak
     gravitational lensing due to density perturbations between
     us and the surface of last scattering).  The red ``scalar''
     curve is the GG power spectrum due to density perturbations
     (with reionization), shown here for reference.  The dotted
     ``lensing'' (green) curve is the
     cosmic-shear contribution to the curl component that comes
     from structures out to a redshift $z=1$, and the green
     ``lensing'' dot-dash curve is the residual cosmic-shear
     power spectrum left after subtraction with higher-order
     temperature-polarization correlations, as described in the
     text.  From Ref. \protect\cite{kck}.}
\label{fig:clbylens}
\end{figure}

In Fig. \ref{fig:clbylens} we show numerical results for the CMB
polarization curl induced by cosmic shear (lensing) \cite{kck,knoxso}.
If the gravitational-wave amplitude (solid line) is large
enough, the CS curl will not interfere with detection of
gravitational waves, and the gravitational-wave signal can be
distinguished from the lensing signal by the shape of the power
spectrum. However, if the gravitational-wave amplitude is
small ($\lesssim 4\times 10^{15}$ GeV), lensing (dashed line)
produces a background, and the gravitational-wave curl cannot be
detected \cite{kck,knoxso,LewChaTur02}.

Fortunately, something can be done to help separate
the CS-induced curl from the GW-induced curl.
Seljak and Zaldarriaga \cite{SelZal99}, Hu and Okamoto
\cite{huoka}, and others have shown that measurement of
higher-order correlations induced by lensing can be used to
reconstruct $\delta\vec{\theta}(\vec{\theta})$,
the displacement, as a function of position on the sky.
Refs. \cite{kck,knoxso} have then evaluated how well these may
be used to reduce the CS-induced curl.  This subtraction is
somewhat involved technically, and still under active
investigation.  We give a very brief description here to provide
the basic flavor of the technique; readers are referred to the
original papers for more details.

In the absence of lensing each Fourier mode of the T (or Q or U)
field is statistically independent:
\begin{equation}
     \VEV{T(\vec{l})T(\vec{l}')} =
     0\:\:\:{\mathrm for}\:\:\:\vec{l}\neq\vec{l}'.
\end{equation}
However, if there is lensing, an observed Fourier mode
$T(\vec{l})$ has contributions from all pairs of temperature and
projected-potential Fourier modes $T(\vec{l}_1)$ and
$\varphi(\vec{l}_2)$ that have
$\vec{l}=\vec{l}_1+\vec{l}_2$. Thus, with lensing,
\begin{equation}
     \VEV{T(\vec{l})T(\vec{l}')}=f(\vec{l},\vec{l}')
     \varphi(\vec{L}) \qquad \vec{l}\neq\vec{l}',
\end{equation}
in the presence of some fixed projected potential
$\varphi(\vec{\theta})$ with Fourier components
$\varphi(\vec{L})$. Here,
\begin{equation}
     f(\vec{l},\vec{l}') = C_l^{\rm TT}(\vec{L} \cdot
     \vec{l})+C^{\rm TT}_l(\vec{L}\cdot \vec{l}').
\end{equation}
(There are analogous expressions for polarization, but for
simplicity, we deal here only with T.)  To determine a given
$\vec{L}$ component of $\varphi$, we simply sum over all pairs
$T(\vec{l}_1)$, $T(\vec{l}_2)$ with
$\vec{L}=\vec{l}_1+\vec{l}_2$. In practice we have to weight
these pairs taking into account noise from the random-field
nature of the fluctuations, as well as instrument noise. Doing
so, Hu-Okamoto find the following estimator for the Fourier
components of the displacement angle,
\begin{equation}
     \delta\vec{\theta}(\vec{L})=
     \frac{i\vec{L}A(L)}{L^2}\int\frac{d^2\vec{l_1}}
     {(2\pi)^2}T(\vec{l}_1)T(\vec{l}_2)F(\vec{l}_1,\vec{l}_2),
\label{eqn:estimatorone}
\end{equation}
\begin{equation}
     F(\vec{l}_1,\vec{l}_2)\equiv\frac{f(\vec{l}_1,\vec{l}_2)}
     {2C_{l_1}^{\rm TT,t}C_{l_2}^{\rm TT,t}},\:\:\:\:\:\:\:A(L)=L^2
     \left[\int\frac{d^2\vec{l_1}}{(2\pi)^2}
     f(\vec{l}_1,\vec{l}_2)F(\vec{l}_1,\vec{l}_2)\right]^{-1},
\label{eqn:estimatortwo}
\end{equation}
and $C_l^{\rm TT,t}$ is the total observed (signal plus noise) power
spectrum.  Thus, with these estimators, the projected potential
can be determined as a function of position across the sky from
the measured temperature map.  We can then use this
projected-potential measurement to reconstruct the polarization
pattern at the surface of last scattering from the (lensed)
polarization pattern that is observed.  Similar estimators that
use the lensed polarization (rather than temperature) can also be
constructed \cite{huoka,kc}, but we do not include them as the
expressions rapidly become unwieldy.  The precision with 
which $\varphi(\vec{L})$ can be reconstructed depends on the
number of small-scale coherence patches in the
temperature-polarization map that can be used as `sources' with
which the shear can be reconstructed.  Thus, high angular
resolution and high sensitivity are required.  Since the
polarization power spectrum
peaks at $l\sim 1000$, rather than $l\sim 200$, there are more
small-scale coherence patches in the polarization than in the
temperature. As a result, a high-sensitivity and high-resolution
polarization map will be required for optimal lensing
reconstruction.  The degree to which the curl component induced
by cosmic shear can be reduced depends on how well this
reconstruction can be accomplished.  For plausible assumptions
about a post-Planck CMB polarization satellite (such as those
that NASA is now considering), the CS-induced curl can be
reduced by a factor $\sim 10$ in power-spectrum amplitude (as
shown in Fig. \ref{fig:clbylens}). This requires a full-sky map
of the temperature and polarization with an angular resolution
$\theta_{\rm fwhm}\sim$arcmins and detector noise-equivalent 
temperature $s\sim1\mu {\mathrm K}\sqrt{{\mathrm sec}}$.
Higher-order correlations may improve upon this subtraction by a
bit more \cite{hirata,amblard}.

\medskip
\noindent {\sl Exercise 13.  Show that the estimators given in
Eqs. (\ref{eqn:estimatorone}) and (\ref{eqn:estimatortwo}) are
the minimum-variance estimators for the projected potential that
can be constructed from the temperature map.  You may
need to go to the original papers to see how this is done.}
\medskip

The best strategy to detect inflationary gravitational waves is
now a subject of some study.  If a curl search probes
inflaton-potential heights $V^{1/4}\gtrsim 4\times10^{15}$ GeV,
then CS-induced shear does not constitute a background. In this
case, one strategy is to integrate deeply on $\sim
5^\circ\times5^\circ$ patch of sky with moderate
$(\sim0.5^\circ)$ resolution.  For a one-year experiment with
NET $s\geq10\mu {\mathrm K}\sqrt{{\mathrm sec}}$ (e.g., Planck, BICEP, QUIET), this is
the best strategy.  However, if the large-angle
temperature-polarization cross-correlation is as large as WMAP
indicates, then there will be a large reionization bump in the
curl power spectrum.  If so, then a similar sensitivity to
gravitational waves might be achievable with a full-sky map
with $\sim0.5^\circ$ angular resolution and similar detector
sensitivity.

However, if the effective detector sensitivity is $s\lesssim10\mu
{\mathrm K}\sqrt{{\mathrm s}}$, then inflaton-potential heights
$V^{1/4}\lesssim 4\times10^{15}$ GeV can be probed. For
these smaller-amplitude gravitational-wave signals, the
CS-induced shear will be a foreground for the $l\sim50-100$
GW-induced $C_l^{\rm CC}$.  This foreground must be removed  with
higher-order correlations, which requires a full-sky map of the
temperature and polarization with $\theta_{\rm
fwhm}\sim$arcmins. This will have to be the route for
post-Planck $s\sim1\mu {\mathrm K}\sqrt{{\mathrm sec}}$
experiment such as CMBPOL. With quadratic estimators, the lowest
detectable inflaton-potential height $V$ is
$V^{1/4}\sim10^{15}$~GeV \cite{kck,knoxso},
although one may be able to do a bit better with higher-order
estimators \cite{hirata,amblard}.  Anything smaller
will be lost in the CS-induced curl, even after subtraction with
higher-order correlations.  However, with a large-angle
reionization bump, as suggested by recent WMAP measurements, the
gravitational-wave signal at large angles may be as easily
detectable as the smaller-angle signal, without being confused
by the CS-induced curl \cite{wayneprivate}.  The ``best'' survey
strategy is still not entirely clear, and will depend on
experimental factors as well as these more theoretical
considerations.  These questions are the subject of several
ongoing NASA mission concept studies.

\section{Closing Comments}

In these lectures, we have provided details of some of the
theory of CMB polarization.  The reader who has successfully
gone through all of the Exercises may take pride in the
knowledge that he/she has achieved a mastery of the
technical aspects of the subject comparable to that of many
researchers in the field.  Of course, these lectures still leave
many essential aspects of CMB polarization theory uncovered.
These include techniques for solving the Boltzmann equations and a
more intuitive understanding of the features of the power
spectra.  The reader should also be aware that the full- or
flat-sky tensor-harmonic formalism must necessarily be altered
to analyze real CMB maps.  Such maps will have sky cuts and so
the measured regions of sky will be less than the full sky and
may be irregularly shaped.  Sophisticated techniques for dealing
with a cut sky have now been developed, and many other issues
that accompany analysis of real maps are highly nontrivial and
completely neglected here.

The past few years have been quite exciting for the CMB.  There
is also clearly a very active foreseeable future in the field
with a number of targets for experimentalists: e.g., the
polarization autocorrelation function, detection of
weak-lensing, and the longer-term goal of detecting inflationary
gravitational-wave background.  If the past is any guide,
however, chances are that the most exciting discoveries are
those that theorists have not yet anticipated.

\nonumsection{Appendix: Rotational invariance of an `spin-2 field'}

Here we demonstrate the rotational invariance of
Eq. (\ref{eq.nabla2}). Let us begin by considering the usual
rotational matrix,
\begin{equation}
\label{eq.lawrot}
     x'_k = A^j_kx_j.
\end{equation}
The derivative operator transforms as
\begin{equation}
\label{eq.partial}
     \frac{\partial F}{\partial x_j} = \frac{\partial
     F}{\partial x'_i}\frac{\partial x'_i}{\partial x_j}.
\end{equation}
From Eq. (\ref{eq.lawrot}), we obtain
\begin{equation}
     \frac{\partial x'_i}{\partial x_j}=\frac{\partial}{\partial
     x_j}\left(A^s_ix_s\right) = A^s_i\frac{\partial
     x_s}{\partial x_j}= A^j_i,
\end{equation}
and Eq. (\ref{eq.partial}) becomes
\begin{equation}
     \frac{\partial F}{\partial x_j} = \frac{\partial
     F}{\partial x'_i}A^j_i.
\end{equation}
Thus, the derivative operator in the rotated system transforms as
\begin{equation}
     \frac{\partial}{\partial x'_i} =
     (A^{-1})^k_i\frac{\partial}{\partial x_k}.
\end{equation}
Taking into account Eq. (\ref{eq.tenstrans}),
Eq. (\ref{eq.nabla2}) becomes
\begin{equation}
\partial'_i\partial'_jP'_{ij} = (A^{-1})^k_i
\partial_k(A^{-1})^l_j \partial_lA^i_mA^j_nP_{mn} =
\delta^l_n\delta^k_m\partial_k\partial_lP_{mn}=\partial_k\partial_lP_{kl}.
\end{equation}

\nonumsection{Acknowledgments}

MK was supported in part by NASA NAG5-11985, and DoE
DE-FG03-92-ER40701. We thank the organizers of the Villa
Mondragone School of Gravitation and Cosmology for inviting
these lectures and for their hospitality.  We also thank
M. Doran, A. Cooray, A. Kosowsky, K. Sigurdson, and especially
J. Pritchard for comments on an earlier draft.

\nonumsection{References}

\end{document}